\documentclass{aastex}
\usepackage{emulateapj5}
\usepackage{psfig}
\usepackage{amsmath}
\usepackage{euscript}

\newcommand{\cyg}{Cygnus X-3}

\newcommand{\msun}{$M_{\odot}$}

\newcommand{\degree}{$^{\circ}$}

\slugcomment{Submitted version of \today}

\shorttitle{Radio Observations of Cygnus X-3 in Flare}
\shortauthors{Miller-Jones et al.\ }

\begin{document}

\twocolumn [\title{Time-sequenced Multi-Radio-Frequency Observations
of Cygnus~X-3 in Flare}

\author{James C.\ A.\ Miller-Jones and Katherine M.\ Blundell}
\affil{University of Oxford, Astrophysics, Keble Road, Oxford, OX1
3RH, U.K.}
\email{jcamj@astro.ox.ac.uk, kmb@astro.ox.ac.uk}
\author{Michael P.\ Rupen and Amy J.\ Mioduszewski}
\email{mrupen@aoc.nrao.edu, amiodusz@aoc.nrao.edu}
\affil{NRAO, Array Operations
Center, 1003 Lopezville Road, Socorro, NM 87801, U.S.A.}
\author{Peter Duffy}
\affil{Department of Mathematical Physics, University College Dublin,
  Dublin 4, Ireland}
\email{peter.duffy@ucd.ie}
\and
\author{Anthony J.\ Beasley} 
\affil{Owens Valley Radio Observatory, Caltech, Big Pine, CA 93513, U.S.A.}
\email{tbeasley@ovro.caltech.edu}

\begin{abstract}
Multifrequency observations from the VLA, VLBA and OVRO Millimeter
Array of a major radio outburst of \cyg\ in 2001 September are
presented, measuring the evolution of the spectrum of the source over
three decades in frequency, over a period of six days.  Following the
peak of the flare, as the intensity declines the high-frequency
spectrum at frequency $\nu$ steepens from $\nu^{-0.4}$ to
$\nu^{-0.6}$, after which the spectral index remains at this latter
terminal value; a trend previously observed but hitherto not
satisfactorily explained.  VLBA observations, for the first time,
track over several days the expansion of a sequence of knots whose
initial diameters are $\sim 8$\,milliarcseconds.  The light-crossing
time within these plasmons is of the same order as the time-scale over
which the spectrum is observed to evolve.  We contend that properly
accounting for light-travel time effects in and between plasmons which
are initially optically thick, but which after expansion become
optically thin, explains the key features of the spectral evolution,
for example the observed timescale.  Using the VLBA images, we have
directly measured for the first time the proper motions of individual
knots, analysis of which shows a two-sided jet whose axis is
precessing.  The best-fit jet speed is $\beta \sim 0.63$ and the
precession period is $\sim 5$ days, significantly lower than fitted
for a previous flare.  Extrapolation of the positions of the knots
measured by the VLBA back to zero-separation shows this to occur
approximately 2.5 days \textit{after} the detection of the rise in
flux density of \cyg.
\end{abstract}

]
\section{Introduction}
\label{sec:intro}

\cyg\ was first detected by \citet{Gia67} using X-ray proportional
counters during a 1966 rocket flight and rose to prominence with the
detection of a giant radio flare in 1972 September \citep{Gre72}.  It
is a high-mass X-ray binary system located in the Galactic plane at a
distance of $\approx$10 kpc from the Earth \citep{Pre00,Dic83}.  With
Galactic co-ordinates $l=79.84$\degree, $b=0.69$\degree, it lies in or
behind either the Perseus or the Outer Arm, depending on its exact
distance from us.  The nature of the compact object is still
uncertain.  The large interstellar extinction to the source precludes
optical spectroscopy, making it difficult to obtain a reliable mass
function and hindering identification of the companion star.  However,
infrared spectra show strong, broad helium emission lines, and a lack
of strong hydrogen emission.  Spectral analysis by \citet{vanKer96}
indicated that the companion is a Wolf-Rayet star of the WN7 subclass,
although more recent observations by \citet{Koc02} suggested that the
subclass may in fact be WN8.  The orbital period, as inferred from
X-ray and infrared flux modulations, is 4.8 hours
\citep[e.g.][]{Par72}.  This source occasionally undergoes huge radio
outbursts where the flux density can increase up to levels of
$\sim20$\,Jy.  Of order 50 outbursts with peak radio flux densities
exceeding 1\,Jy have been observed since the detection of the first
such flare in 1972 September.  During one of these radio outbursts,
the hard X-ray emission was found to correlate strongly with the radio
emission, although the soft X-ray emission was anticorrelated with
radio flux density \citep{McC99}.  In quiescence, the soft-X-ray is
found to be strongly correlated with the radio emission, and
anticorrelated with the hard X-rays \citep{Cho02}.  This
soft-X-ray---radio correlation disappears immediately before and
during periods of radio flaring \citep{Wat94}.  During recent
outbursts, jet-like structures have been observed at radio
frequencies, although there has been some debate as to the morphology
of the collimated emission.  Two-sided jets from a flare in 2000
September were detected on arcsecond scales in a north-south
orientation with the Very Large Array (VLA) \citep{Mar01}, with an
inferred jet speed $v = 0.48c$. On milliarcsecond scales, a one-sided
jet with the same orientation was imaged with the Very Long Baseline
Array (VLBA) after a flare in 1997 February \citep{Mio01}.  In this
case, the derived jet speed was significantly higher ($v \geq 0.81c$).

In this paper we analyse radio observations of the 2001 September
outburst of Cygnus X-3 and describe some of the physical mechanisms
responsible for the observed evolution of the source.  In
\S\ref{sec:prev_work} we summarize the characteristics of previous
radio flares of \cyg.  In \S\ref{sec:VLA}, \S\ref{sec:ovro} and
\S\ref{sec:vlba}, we present our observations from the VLA, Owens
Valley Radio Observatory (OVRO) Millimeter Array, and the VLBA
respectively, and in \S\ref{sec:vlba_analysis} we analyse the VLBA
images and infer the speed, sidedness and precession parameters of the
jet.  We go on to discuss the spectral evolution and the underlying
physical mechanisms for the low-frequency turnover in
\S\ref{sec:spectra}.  We then derive constraints on the magnetic field
strength, electron number density and energy density in the source in
\S\ref{sec:src_parms}, and describe the physics behind the
high-frequency spectral evolution in \S\ref{sec:analysis}.  In
\S\ref{sec:big_guys} the magnetic field strength and energy in the jet
in this outburst of \cyg\ are compared to derived parameters of radio
galaxies such as Cygnus\,A and M\,87.

\section{Spectral characteristics}\label{sec:prev_work}

The temporal evolution of the radio spectrum of \cyg\ was first
studied during the outburst of 1972 September \citep[][and 20 papers
following this]{Gre72}.  This outburst was observed between 0.4\,GHz
and 90\,GHz, and observations were made during both the increase and
the subsequent decrease of the radio flux density.  This outburst of
\cyg\ exhibited approximately similar characteristics to several other
major flares observed from this system: a short-lived exponential
fading of the intensity \citep[e.g.][]{All72, Hje74, Mar74, Wal95}
with $e$-folding times $\tau$ in the range $0.15<\tau<2.75$\,days,
followed by a power-law decay in intensity.  With sufficient frequency
coverage a two-phase spectral evolution is seen; an initial steepening
of the spectrum, after which the spectral index $\alpha$ (which
relates the intensity $S_{\nu}$ to the frequency $\nu$ as
$S_{\nu}\propto \nu^{-\alpha}$) remains at a terminal value of
$\sim0.6-0.7$ \citep[e.g.][]{Gre72b, All72, Hje72, Sea74, Gel83,
Fen97}.  In the dataset we present in this paper, we have observations
over a wider frequency range than made previously, which more tightly
constrain the observed evolution.  However, our observations were not
triggered early enough to observe the rise phase of the outburst,
which precludes detailed modelling of the mechanism responsible for
launching the jet, and comparison with previous models purporting to
explain the rise phase of radio flares \citep[e.g.][]{Mar92}.

\section{VLA observations and data reduction}\label{sec:VLA}

The outburst of \cyg\ was observed to begin when the radio flux
density started to rise on 2001 September 14 (MJD $52167\pm0.2$)
\citep{Tru02}, and was monitored regularly with the VLA from September
18 until ten days following the outburst, and then less frequently
until the end of October.  The VLA was in its most compact ``D''
configuration, so that the source was unresolved and, crucially, the
VLA measured the integral of all emission regions associated with the
outburst.  The source was observed using snapshot observations of
duration two to three minutes, sufficient to detect and perform
photometry on the source.

Data were taken with the VLA using standard procedures, observing at
eight different frequency bands (centred on 73.8\,MHz, 327.5\,MHz,
1.425\,GHz, 4.86\,GHz, 8.46\,GHz, 14.94\,GHz, 22.46\,GHz, and
43.34\,GHz).  A summary of the observing dates and frequencies may be
found in Table \ref{tab:vla_observations}.  All observations except
for those at 73.8-MHz were made in two independent frequency bands (IF
pairs).  All IFs were of width 50\,MHz except for the 327.5- and 73.8-MHz
data, which had bandwidths of 3.125 and 1.563\,MHz respectively.  At
1.425\,GHz, the two IFs (with frequencies of 1.465 and 1.385\,GHz for
the upper and lower sidebands respectively) were imaged separately, to
gain an extra low-frequency point where the spectral shape was
changing most rapidly.  At all other frequencies, the data in the
separate IFs were averaged together during imaging.  Calibration and
image processing were performed using NRAO's Astronomical Image
Processing System (\textit{AIPS}).  The primary calibrators used were
3C\,48 and 3C\,286, depending on which was above the horizon at the
time of the observation.  One of these was observed at each frequency
in every observing run.  The flux scale used was that derived at the
VLA in 1999, as implemented in the 31Dec02 version of \textit{AIPS}.
Several different secondary calibrators were used, depending on the
frequency of observation and on the other sources observed during the
run.  These were J\,2007+4029, J\,2015+3710, and J\,1953+3537, at
angular separations of 4.7$\degr$, 5.0$\degr$, and 9.3$\degr$ from the
source respectively.  After initial calibration, the data were imaged
and self-calibrated (initially using phase-only self-calibration, but
ultimately using simultaneous amplitude and phase self-calibration to
make the best possible images) in order that photometry could be
performed on the source.

\subsection{Low frequency issues}

At low frequencies, the large antenna primary beam required a slightly
different observing strategy.  The powerful radio galaxy Cygnus\,A is
located only 6.3$\degr$ away from \cyg\, and, at 17\,kJy, is the
brightest source in the 74-MHz sky.  At 330\,MHz, the flux density of
Cygnus\,A is 5.3\,kJy.  Different methods were employed at the two
different frequencies.  At 74\,MHz, where the field of view is
largest, a single observation was taken, with the pointing centre
located halfway between Cygnus A and \cyg.  At 330\,MHz, alternate
scans were made on Cygnus A and on \cyg\, as for the higher-frequency
observations.  It was thought that the former method might allow the
deconvolution process to remove the sidelobes from Cygnus A more
accurately.  The results suggested the latter method to be preferable,
since a source was detected at the known position of \cyg\ at
330\,MHz.  At 74\,MHz however, it was only possible to put an upper
limit on the flux density of the source, taken as three times the
r.m.s.\ background noise level of the final sky image at the location
of \cyg.

\subsection{High-frequency issues}
\subsubsection{Opacity corrections}

At the highest frequencies (43\,GHz and in some cases, 22\,GHz), the
weather had a significant effect on the data, causing poor phase
correlation.  This required scalar averaging or vector averaging over
a short timescale (of order the integration time of 3.3\,s) of the
visibilities during the calibration process, rather than scan-based
vector averaging as is customary.  In addition, the variable
atmospheric opacity required the use of weather (WX) tables or tipping
scans (in which the system temperature for an individual antenna is
measured as the antenna slews from the zenith to the horizon; the
dependence of system temperature on zenith angle may be used to
determine the opacity) in order to correct for the effects of the
opacity, and thus determine the flux scale accurately.  Tipping scans
were only available for the 22-GHz observations made on September 18
and 19.  For all other epochs and frequencies, an estimate of the
opacity had to be made using the weather information taken during the
observations.  The fitted opacity curves given by \citet{But01} were
initially used for this.  However, more recent analysis by
\citet{But02} showed that an improved estimate of the atmospheric
opacity may be made by including a term to account for seasonal
variations, which we therefore factored into the calculations.  The
use of a constant estimated opacity throughout an individual
observation was justified by the short lengths of the observations.

\subsubsection{Referenced pointing}

Another problem in making high-frequency observations is the pointing
accuracy of the antennas: the difference between the centre of the
primary beam and the desired source position.  Factors which may
contribute to the pointing error include atmospheric refraction,
gravitational deformation and differential heating of the antennas and
non-perpendicularity or misalignment of the antenna axes
\citep{Cla73}.  Under normal observing conditions, the VLA pointing is
accurate to 10-20\hbox{$^{\prime\prime}$}, but can be as bad as an
arcminute in poor weather conditions \citep{Rup97}.  At high
frequencies, the primary beam is small (the full-width at half power
being about 1\hbox{$^\prime$}\ at 43\,GHz and 2\hbox{$^\prime$}\ at
22\,GHz).  Thus the pointing error can be a large fraction of the
primary beam, leading to an error in the observed amplitude of the
source.  To account for this, primary referenced pointing was used in
the observations (deriving the pointing offsets by pointing up on the
calibrator at a lower frequency --- typically 8.4\,GHz --- and
applying the derived corrections).  This lowers the r.m.s.\ pointing
error to about 2\hbox{$^{\prime\prime}$}\ in azimuth and
5\hbox{$^{\prime\prime}$}\ in elevation at elevations below about
80\degree.  Wind loading of the antennas may also affect the pointing,
but the effect of winds below 6\,ms$^{-1}$ has been measured to be
insignificant \citep{Rup97}.  Fortunately, the maximum wind speed in
any of our observations was $5.7$\,ms$^{-1}$.  For the observing run
early on September 21, the pointing accuracy in azimuth for a subset
of the VLA antennas was poor at elevations $\gtrsim 80$\degree,
leading to significant gain fluctuations.  After removing the affected
data, we found that the point-source flux densities before and after
amplitude self-calibration agreed to within 1\%.

\subsubsection{Gain corrections}

As the VLA antennas track in elevation, the variation in gravitational
deformation of the antennas causes the reflector profiles to change.
This makes a difference to the response of the telescope at high
frequencies (at and above 15\,GHz, where the wavelength is more
comparable to the scale of the distortions).  To account for this
measurable effect, the gain curves closest in time to the observing
dates (those of 2001 November), as measured by NRAO staff, were used.

\subsection{Errors and Uncertainties}\label{sec:errors}

There are several possible sources of uncertainty in the data
reduction process which must be taken into account when estimating the
uncertainties on the VLA measurements.  The r.m.s.\ noise in the final
image of the source gives an idea of the random error in the
measurements due to thermal noise fluctuations.  For 2-minute
integrations with the full VLA, the expected thermal noise is at best
0.1\,mJy per beam at 8.4\,GHz.  At higher frequencies, the system
temperature increases due to atmospheric emission, so the noise rises
to 0.6 mJy per beam at 43\,GHz.  At low frequencies, the Galactic and
cosmic backgrounds increase the system temperature markedly, to 34 mJy
per beam at 330\,MHz and 3.3 Jy per beam at 74\,MHz for observations
in the Galactic plane.  In practice, even this sensitivity level will
not be reached for low-frequency snapshot imaging, since in this case
it is impossible to solve accurately for the sidelobes of confusing
sources with only a short observation.

In addition to the thermal noise in the image, there are several
sources of uncertainty resulting from the calibration process, the
most significant of which is likely to be the uncertainty in the
determination of the flux scale.  The \textit{AIPS} tasks
SETJY and GETJY quote nominal uncertainties in the
calibrator fluxes, but these are based solely on the scatter in the
amplitude gain ratios.  If there are other effects, such as phase
decorrelation of a weak phase calibrator (such as J\,1953+3537 at high
frequencies, which was used occasionally) or due to bad weather, then
depending on the form of averaging and the solution interval used in
the \textit{AIPS} task CALIB, the derived antenna voltage
gain solutions and hence the flux scale will vary.

CALIB derives the complex antenna voltage gain solutions for amplitude
and phase for sources with known flux densities and structures, thus
setting the amplitude scale.  To reduce the error in the derived
gains, the calibrator data are averaged in time.  To obtain an
unbiased estimate of the phase, vector averaging is used.  However, it
is more difficult to obtain an unbiased estimate of the amplitude,
since vector averaging will potentially underestimate the amplitude
due to decorrelation by phase noise.  Alternatively, scalar averaging
will tend to increase the amplitude due to the inclusion of a noise
bias; a scalar average of a data set consisting of pure noise will
produce a positive signal.  However, this is only a serious problem
for high-frequency observations, where the phase decorrelation tends
to be greater, and for weak ($\lesssim 5\sigma$) sources.  During the
observations in September, \cyg\ was bright ($\approx 10^2-10^3
\sigma$), so scalar averaging should not introduce a significant noise
bias.  Since the amplitude and phase fluctuations of most instrumental
effects are uncorrelated \citep{Fom99}, separate amplitude and phase
averaging may be used to give the most accurate estimates of each
(although when deriving the actual antenna-based solutions for the
complex gains, the noise bias is reduced by solving simultaneously for
both).  In the data reduction process, the initial approach was to use
vector averaging over a whole scan, solving for amplitude and phase
simultaneously.  At high frequencies ($>15$\,GHz) scalar averaging was
then also done as a check.  If the two disagreed, separate calibration
was done for phase and amplitude, solving for phase first using vector
averaging, and then applying those gains and solving simultaneously
for amplitude and phase using scalar averaging.

According to the calibration procedure laid out in the VLA calibration
manual, the flux density bootstrapping is accurate to 1-2\% for the
1.4-, 4.8-, and 8.4-GHz observing bands.  The uncertainty in the
measurements was taken as 1\% for 8.4\,GHz and 2\% for 1.4 and
4.8\,GHz.  At higher observing frequencies, the accuracy is possibly
as good as 3-5\%, but effects such as poor antenna pointing and
atmospheric opacity are thought to reduce the accuracy further.  From
the data reduction, it was clear that the atmosphere was particularly
troublesome on certain days (September 18, 21, 23, and 24), and phase
decorrelation made the flux scale uncertain.  On those days, the
percentage uncertainty was taken as the percentage flux density
difference between the GETJY flux densities of the phase calibrators
found using vector averaging and scalar averaging during the
calibration.  This difference was only found to be significant at
43\,GHz.  For the better days (September 19 and 22), and for the
observations at 15 and 22\,GHz, the uncertainty was derived by
spectral fitting.  These uncertainties were then combined in
quadrature with the thermal noise in the images (although since we
achieved a dynamic range of several hundred to one in most of the
images, the contribution of the thermal noise to the total uncertainty
was minimal).

It seems reasonable to assume that the spectrum is smooth.  Fitting a
power law to the spectrum at frequencies well above the low-frequency
turnover, the deviations of the high-frequency points from the expected
smooth curve were measured.  The uncertainty in the measurement for a
particular frequency was then taken as the r.m.s. deviation of the
points from their respective power laws (taking the mean over all six
days).  This resulted in uncertainties of 4.8\% for the 15-GHz data,
7.7\% for the 22-GHz data, and 8.0\% for the 43-GHz points not
affected by phase decorrelation (for the data so affected, the
uncertainty was of order 7-12\%).

At low frequencies (74\,MHz and 330\,MHz), the flux scale was
determined from observations of Cygnus\,A.  It is thought to be good
to 5-10\% (R.A. Perley 2003, private communication).

\section{OVRO observations and data reduction}\label{sec:ovro}
Five observations of \cyg\ were made by the OVRO Millimeter Array
during the period September 20 to 26 (see
Table\,\ref{tab:ovma_observations}), each involving between 1 and 3
hours of integration.  This array is comprised of six 10.4\,m antennas
on a 400\,m T-shaped track, situated at 1220\,m altitude in the high
desert of eastern California.  The array was in compact configuration
during these observations, and had a typical angular resolution of
$\sim11$\,\hbox{$^{\prime\prime}$}$\times
7.5$\,\hbox{$^{\prime\prime}$}. Two 1-GHz sidebands centred at 97.98
and 100.98\,GHz were observed at a single polarisation at all epochs.
The flux density scale was set by assuming the flux density of the
adjacent continuum calibrator J\,2015+3710 was 2.87\,Jy for all epochs
(this value was derived from the first observations made on September
20 involving Uranus and 3C\,345, and incorporates the on-line opacity
correction), and is considered accurate at the 10\% level. Reduction
of the data was performed using the NRAO \textit{AIPS} package.  The
typical 1$\sigma$ image detection limit for the observations was
5-10\,mJy.  The lightcurves obtained using the VLA and OVRO Millimeter
Array are shown in Figure\,\ref{fig:vla_lightcurves}.

\section{VLBA Observations and data reduction}\label{sec:vlba}
\begin{figure}[t]
\psfig{figure=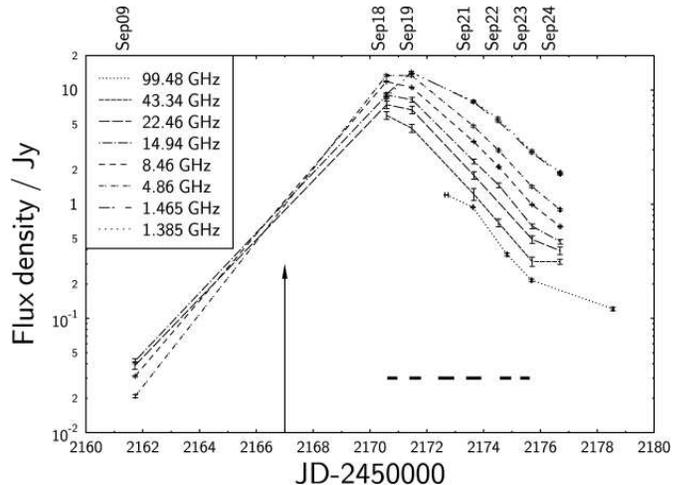,width=8.8cm,angle=0,clip=t}
\caption{VLA and OVRO Millimeter Array observations of the integrated
  flux density during the outburst of \cyg.  The bold black horizontal
  lines indicate the times of the VLBA observations.  The arrow
  indicates the start of the outburst, as detected with the RATAN-600
  telescope \citep{Tru02}.  Lines have been drawn between the points
  corresponding to September 9 and September 18, to indicate the
  change in spectral shape.  This does not represent the rate at which
  the flux density rose during the outburst, as there is insufficient
  temporal coverage to accurately plot the rise phase of the flare.
  Note how the spectrum goes from rising with frequency before the
  flare to falling with frequency afterwards, a trend previously
  observed as the source moves from the pre-flare ``quench'' state to
  outburst \citep[e.g.][]{Fen97}.  A power-law fit to the high
  frequency decay curves between September 19 and September 23 yields
  a power-law index of $3.8\pm0.2$. \label{fig:vla_lightcurves}}
\end{figure}

\cyg\ was also monitored with the VLBA during the September outburst.
Each day from 2001 September 18 to 23, observations of duration
between 8 and 13 hours were made using all ten VLBA antennas and a
single dish of the VLA.  The observation periods are shown in Figure
\ref{fig:vla_lightcurves}.  On all six days, observations were made at
1.660, 4.995, 15.365, and 22.233\,GHz, with the time being split
approximately equally between the different frequencies.  Scans at a
given frequency were spread over the whole observation period in order
to improve the \textit{uv}-coverage.

The VLBA observations were carried out using dual polarisation and
two-bit sampling.  At all four frequencies, an 8-MHz bandwidth was
used.  The two IF pairs were allocated across an uninterrupted
frequency range, rather than being more widely spaced.  Scans on \cyg\
were interleaved with scans on a phase calibrator source, with a
3-minute cycle time (70\,s on the calibrator, 110\,s on the source).
The calibrators used were J\,2052+3635 (5.9\degree\ from the source)
for September 18 and 19, and J\,2007+4029 (4.7\degree\ from the source)
for all other epochs.  The data were correlated using the VLBA
correlator in Socorro, New Mexico.  All data taken at $<$15\degree\
elevation were flagged.  Amplitude calibration, fringe-fitting and
imaging were carried out using the \textit{AIPS} software according to
NRAO guidelines laid out in Appendix C of the \textit{AIPS} Cookbook.
The flux density scale was set using system temperatures and antenna
gains, and is believed to be accurate to $\sim5\%$ according to the
VLBA Observational Status Summary.  Fringe-fitting was carried out
only on the calibrator sources, not the source itself, except at
22\,GHz as described later in this section.

Unfortunately, \cyg\ is one of the most heavily scatter-broadened
sources in the radio sky owing to its location in the direction of the
Cygnus OB2 association \citep{Wil94,Mio01}.  The size of the
scattering disk scales as $\nu^{-2}$, so particularly at the lower
observing frequencies, much of the information from the long baselines
was lost, and the resulting \textit{uv}-coverage of the observations
was poor.  In addition, the images of the phase calibrators themselves
were heavily scattered, so it was not possible to calibrate the outer
antennas at low frequencies.  Only the baselines between the antennas
in the southwestern United States (the VLA antenna, Pie Town, Los
Alamos and Fort Davies) yielded useful information about the source.
This meant that we were unable to produce any good quality images at
1.6\,GHz, and that the resolution decreased more markedly with
frequency than would have been expected from a simple $\lambda/D$
decline.

In addition to sparse \textit{uv}-coverage (the VLBA itself has a
maximum of ten antennas, the outermost of which were lost owing to the
scattering of information from long baselines), the complex, variable
nature of the source during each observation posed further
difficulties for the data reduction.  In an observation of duration 13
hours, the components of the source would not only change in flux
density, but would also gradually change position, owing to the motion
of the jet during the period of observation.  Modelling the latter
effect however revealed that the positional changes would simply
elongate the components in the images.  This effect would in fact only
be a problem at the highest frequency, since for a 13-hour
observation, a proper motion of $\sim 10$\,milliarcseconds (mas) per
day would shift the position by 6.5\,mas.  This is a little smaller
than the beam size at 22\,GHz (see Table \ref{tab:vlba}),
significantly less than the beam size at 15\,GHz and much smaller than
the size of the scattering disk at 5\,GHz.  So even at 22\,GHz, the
effect would be fairly small.  The earlier epochs of VLBA data
suffered most from source variability, whereas the relatively low
surface brightness of the source caused most of the problems during
the later observations.  On September 22 and 23, we were unable to
produce any good images.

\begin{figure}[t]
\centering{
\psfig{figure=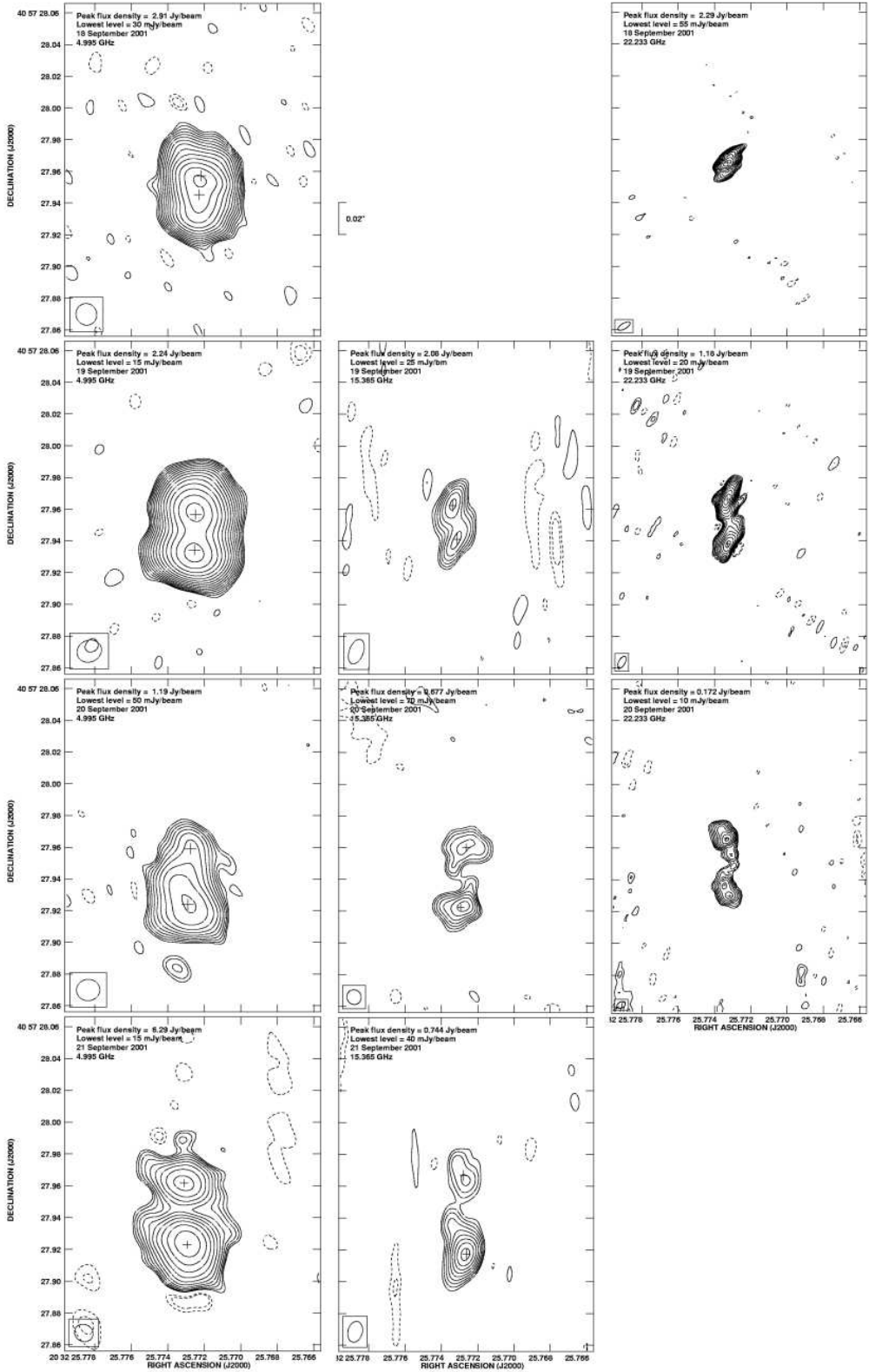,width=14cm,angle=0,clip=t}
}
\caption{VLBA images of Cygnus X-3 in flare.  Images are shown at
  4.995\,GHz in the left column, 15.365\,GHz in the central column,
  and 22.233\,GHz in the right column.  Where images are missing, the
  data were not good enough to make a good quality image.  Time
  progresses down each column, each row a day later than the previous
  one.  Solid contour levels are $(\sqrt{2})^n$ times the lowest level
  stated in the images ($n=0,1,2,\ldots$).  Dashed contours are $-1$
  and $-\sqrt{2}$ times the stated lowest levels.  The beam sizes are
  shown in the lower left corner of the images.  Crosses indicate the
  fitted positions of the maxima, as plotted in
  Figure\,\ref{fig:vlba_positions}.  Assuming a distance of 10\,kpc to
  the source, 1\,mas on the images corresponds to a projected
  separation of 10\,AU.
  \label{fig:vlba_images}}
\end{figure}

Several approaches to producing good quality images of the source were
tried.  Splitting the observation into small timeranges and imaging
each individually in order to reduce the effects of variability and
proper motion was found to help for the early observations when the
source was bright, but for the later epochs, did not leave enough flux
density to make a good quality image of the source.  Self-calibration
with a point source model was attempted, but this was not found to
improve the image significantly.  Imaging only the part of the
observation where a plot of the amplitude of the short baselines
against time was relatively flat, in order to reduce the effects of
variability, was again not found to improve the image.  Downweighting
the central regions of the \textit{uv}-plane (corresponding to the
short baselines) by the use of super-uniform weighting was similarly
found to be of limited use, as was the attempt to model the
visibilities with Gaussian components.  The method that did most to
improve the map was to image and self-calibrate the phase calibrator
(J\,2007+4029), and then to transfer the solutions for amplitudes and
phases to the target.  Once the solutions had been transferred, some
of the methods listed above were applied once more in an attempt to
improve the images.  Of these, only downweighting the short baselines
was found to significantly improve the images.  The resulting images
are shown in Figure\,\ref{fig:vlba_images}.  More information on the
images (beam sizes, noise levels and exact observation times) is given
in Table \ref{tab:vlba}.  Only phase self-calibration was used since
amplitude self-calibration was not found to improve the images.

The observations at 5 and 15\,GHz were phase-referenced to a
calibrator (J\,2052+3635 or J\,2007+4029 as detailed above) in order
to preserve positional information.  Since self-calibration was
carried out on both the phase calibrator and \cyg\ itself, there might
still have been some degree of uncertainty in the absolute sky
positions of the images.  In order to estimate this effect, a check
source was imaged, both phase-referenced to the phase calibrator and
to itself.  The difference in the position of the check source in the
two images gave an estimate of the positional uncertainty introduced
in the imaging and self-calibration processes.  The check sources used
were J\,2025+3343 for September 18 and 19 and J\,2052+3635 for all
other observations.

At 22\,GHz, the phase calibrator sources were insufficiently bright,
and the images could not be phase-referenced.  Thus their absolute sky
positions are not accurate; an inevitable consequence of the fringing
process is that the brightest point in the source becomes the centre
of the image.  At 15\,GHz and 5\,GHz however, phase referencing was
carried out.

\onecolumn

\twocolumn

\section{VLBA image analysis} \label{sec:vlba_analysis}

The linear alignment of features in the VLBA images are interpreted as
jets in an almost north-south orientation, in agreement with previous
work \citep[e.g.][]{Mio01, Mar01}.

\begin{figure}
\psfig{figure=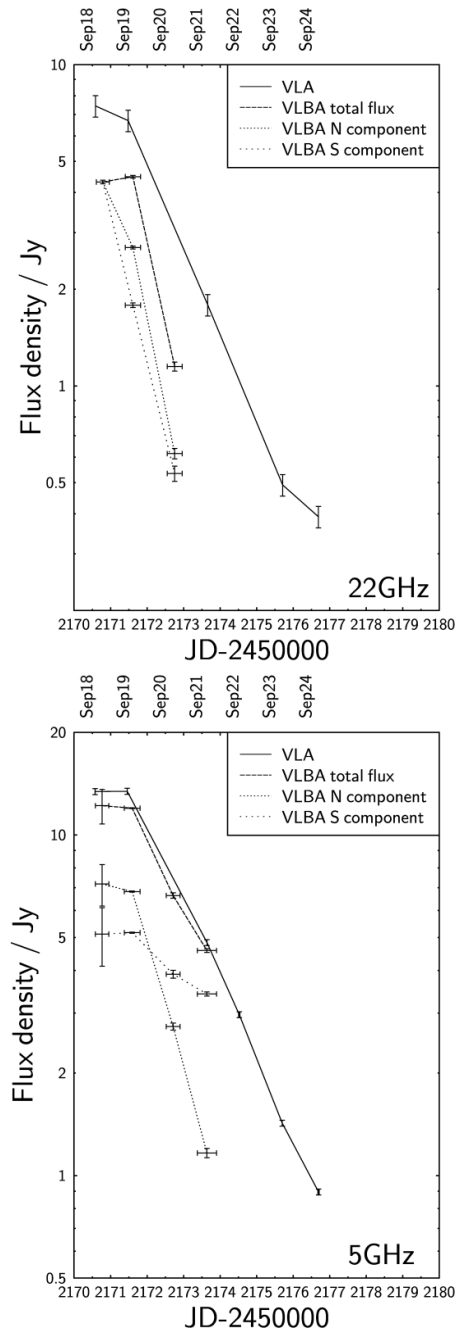,width=6cm,angle=0,clip=t}
\caption{22-GHz (top) and 5-GHz (bottom) VLBA component fluxes.  The
  total flux detected with the VLBA tracks the flux detected with the
  VLA fairly accurately.  Note how the northern component flux
  density drops dramatically at both frequencies between September 19
  and September 20, a decrease mirrored by the southern component at
  22\,GHz, but not at 5\,GHz.  The northern and southern components
  could not be distinguished at 22\,GHz on September 18, and the flux
  density of the single measured component has been plotted for both,
  as it was not known how the flux density was split between them.
  \label{fig:vlba_fluxes}}
\end{figure}

The total flux densities recovered from the images track fairly well
the flux densities measured with the VLA
(Figure\,\ref{fig:vlba_fluxes}).  Perhaps a few hundred mJy is missing
at 5\,GHz (i.e.\ $\sim 5$\,\%), although up to a few Jy (corresponding
to several tens of percent) is not recovered at 22\,GHz.  However, it
was difficult to measure accurately the flux densities of the
different components from the images, mainly because of insufficient
resolution.  Hence it was difficult, especially for the images in
which the r.m.s.\ noise is a significant fraction of the component
flux density, to derive accurate spectral indices for the emission.
At 22\,GHz, the northern and southern components (initially
unresolved) fade at similar rates, whereas at 5\,GHz, the northern
component fades more rapidly than the southern component.

When considering the images, care should be taken not to overinterpret
the details of the structures they show.  In the 5-GHz and 15-GHz
images, only the gross features should be relied upon, owing to the
size of the scattering disk (16\,mas at 5\,GHz) or the high r.m.s.\
noise of the images (126\,mJy per beam in the 15-GHz image of
September 19).  The fine structure in the 22-GHz images is certainly
believable, owing to the low r.m.s.\ noise levels achieved and the
high resolution of the images (a small beam size and a small
scattering disk).  Certainly on September 20, the shape of the jet,
the extension to the north of the northern maximum, and the double
peak to the south are believed to be real.

\subsection{Overall morphology}
Accepting that the extended emission we observed with the VLBA
constitutes a jet, we now consider the nature of the jet; whether it
consists of a single pair of ejecta, a continuous set of discrete
ejecta, or a continuous hydrodynamic flow.

The lower-frequency and hence lower-resolution (5-GHz and 15-GHz)
images show the source to be dominated by two bright components.
However, the highest-resolution image at 22\,GHz reveals a sequence of
several discrete knots, with the flux density being dominated by a
single northern component (which we identify in \S\ref{sec:core} and
\S\ref{sec:morphology} as the core with a northern extension) and a
southern component which itself shows hints of being resolved into two
distinct peaks.  It seems most likely, therefore, that the jet
consists of a series of discrete knots, but is dominated by two
brighter components; one in the north and one in the south.  In most
of our subsequent analysis and modelling we concentrate on only the
brightest observed components seen at 5 and 15\,GHz, since they appear
to dominate the flux density of the source and since we believe them
to be uncontaminated by core emission (see
\S\,\ref{sec:vlba_spectra}).  Figure\,\ref{fig:vlba_positions} shows
the movement of the brightest emitting regions with time.

Assuming the likely nature of the jet as a whole series of discrete
ejecta, following the motions of the brightest components can help to
determine some of the fundamental jet parameters, e.g.\ the jet speed
(\S\,\ref{sec:proper_motion}).  Measuring the evolution of the sizes
of the brightest 22-GHz knots constrains the expansion speed of the
plasmons and is evidence that we see discrete blobs.  However, by
considering the underlying nature of the jet, the positions of all the
discrete knots seen in the 22-GHz image from September 20 can be
fitted to give information on the way the jet precesses
(\S\,\ref{sec:morphology}).

\subsection{Jet speed} \label{sec:proper_motion}
\begin{figure}
\psfig{figure=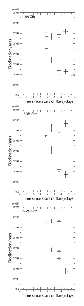,width=4.7cm,angle=0,clip=t}
\caption{Movement in Declination of the VLBA knots with time.  Only
Declination is plotted here because the jet axis is very close to
north-south \citep[\S\,\ref{sec:morphology} and][]{Mio01} and the
knots are expected to move ballistically out from the core.  Positions
were measured by fitting one or more Gaussians to the emitting regions
using the \textit{AIPS} task JMFIT and finding the positions of the
fitted peaks.  Fitted positions are shown with crosses in
Figure\,\ref{fig:vlba_images}.  For the points without a corresponding
image in Figure\,\ref{fig:vlba_images}, the position is taken from the
obvious brightest point in a phase-referenced image which had not been
self-calibrated.  These images had not been included in
Figure\,\ref{fig:vlba_images} as we were unable to produce
good-quality images for these data sets.  Uncertainties in the fitted
positions are taken as the canonical value of 25\% of the beamwidth.
Uncertainties in the time are taken to span the course of each
observation.  Start of flare is taken as JD $2452167\pm0.2$
\citep{Tru02}, i.e.\ 2001 September 14, but see
\S\ref{sec:flare_start} for discussion of when the ejection event
occurred. \label{fig:vlba_positions}}
\end{figure}

The 15-GHz and 5-GHz data were phase-referenced, and analysis of the
positions (depicted in Figure\,\ref{fig:vlba_positions}) of the two
most prominent peaks in the VLBA images suggested that both the
northern and southern components had non-zero proper motions.  (We
chose to ignore the 22-GHz data in these fits because these were not
phase-referenced i.e.\ did not have absolute positional information
and, for the reasons described in \S\ref{sec:vlba_spectra}, because at
this high frequency we believe there to be a significant core
contribution.)  We fitted a straight line to the measured positions at
5\,GHz and 15\,GHz (both data sets taken together) as plotted in
Figure\,\ref{fig:vlba_positions} (top and centre).  We first
constrained the gradient of the northern line to be zero (i.e.\ a
horizontal line, indicating no systematic motion), which fit had a
reduced $\chi^2$ of 3.8.  Fitting for an unconstrained gradient gave a
significantly lower reduced $\chi^2$ of 0.64, with a best fit value
for the proper motion of $2.8\pm0.5$\,mas per day.  This is smaller
than the best-fit value for the southern component ($8.4\pm1.7$\,mas
per day).  Assuming the jet is two-sided (which we substantiate in
\S\,\ref{sec:sidedness}), an analysis of the proper motions of the
approaching and receding jets \citep[e.g.][]{Mir99} determines
$\beta\cos i=0.50\pm0.10$ (where $\beta=v/c$ and $v$ is the
\textit{intrinsic} speed at which the knots move away from the core,
and $i$ is the inclination angle of the jet to the line of sight).
This analysis also gives an upper-limit to the distance to the source
of $d=35.7\pm4.8$\,kpc and an ensuing upper limit to the inclination
angle of the jet axis to the line of sight of
$i_{\rm{max}}=59.8\pm6.7$\degree.  Assuming a source distance of
exactly 10\,kpc, we can solve for the jet speed and inclination angle
to get $\beta=0.56\pm0.09$ and $i=25.7\pm6.5$\degree.

A further check on derived proper motions is often made using a
Doppler boosting analysis.  Given the fading of the intensities of the
plasmons with time (e.g.\ Figure\,\ref{fig:vlba_fluxes} and
\S\,\ref{sec:ltt}) a simple Doppler boosting analysis is not valid.  A
refined application of this method \citep{Mir99} which involves
measuring the flux density in the approaching and receding knots at
equal angular separations from the core is in this case inapplicable
because of the pronounced side-to-side asymmetry which is predicted
from the jet speed and inclination angle derived above and which is
seen in Figure\,\ref{fig:precession}.

\subsection{Plasmon size} \label{sec:plasmon_size}
From the observations, it was possible to derive a maximum size to the
source components, which can be used to help estimate the magnetic
field in the emitting region.  Since the images of \cyg\ are highly
scatter-broadened \citep{Wil94}, particularly at low frequencies, the
apparent source size is dependent on frequency.  \citet{Mio01}
reported that the apparent source size scales as
\begin{equation}
\theta=448(\nu/1\,\rm{GHz})^{-2.09}\quad\rm{mas}.
\end{equation}
which, within the errors, they found to be consistent with a
$\nu^{-2}$ scaling.  Thus to determine the true source size, and to
obtain the best resolution, we used our highest-frequency
(22\,GHz) images.  The \textit{AIPS} task JMFIT was used to fit
Gaussians to the two brightest knots in the images, and the
deconvolved sizes and orientations are shown in
Table\,\ref{tab:vlba_sizes}.  The quoted sizes do not include a
deconvolution from the scattering disk, the size of which is given for
the different observing frequencies in Table\,\ref{tab:vlba}.  The
observed plasmon sizes increased with time, from $7.9_{-3.0}^{+0.5}$
to $16.9_{-3.6}^{+0.9}$\,mas; the errors on these sizes include the
apparent elongation due to proper motion of the knots over the course
of each observation.  For the 8.5-hour observation of September 18,
and a derived proper motion of 8.4\,mas\,d$^{-1}$, the possible
elongation of the knots is of order 3.0\,mas.  Thus the intrinsic size
could be as small as 4.9\,mas.  What may complicate this simple
measurement of the the plasmon size however, is the possibility of
there being multiple plasmons, as observed in the September 20 image,
which sequentially make a contribution to the observed flux density,
but for which the VLBA has insufficient angular resolution.  Assuming
that the measurements (incorporating the uncertainty due to the
smearing effect of component proper motion) do reflect the plasmon
expansion, the expansion speed is then $4.6^{+0.8}_{-2.5}$\,mas per
day, corresponding to a speed of $0.27^{+0.05}_{-0.14}\,c$ for an
assumed source distance of 10\,kpc.  This is, reassuringly, fairly
consistent with the plasmon expansion speed of $0.1-0.3c$ fitted by
\citet{Ato99} for the 1994 outburst of GRS\,1915+105.

\subsection{Is the jet one-sided or two sided?}
\label{sec:sidedness}
There is some discussion in the literature as to whether the observed
jet in \cyg\ is one-sided (owing to Doppler boosting effects reducing
the flux density of the receding jet below the map noise as in
\citet{Mio01}) or two-sided \citep[e.g.][]{Mar01}.  In the following
sections, we will try to assess whether the VLBA images of
Figure\,\ref{fig:vlba_images} show evidence for a one-sided or a
two-sided jet.

\subsubsection{Evidence from proper motions}
From the proper motion analysis of the 5-GHz and 15-GHz data in
\S\ref{sec:proper_motion}, the fits seem to favour the northern
component seen at these frequencies being a moving jet rather than a
stationary core.  This argues for the observed VLBA jet being
two-sided.  This conclusion should be taken with the caveat that
although the images were phase-referenced, they were also
self-calibrated, a process which may in principle move component
positions around by a fraction of a beam.  Given that the beam size at
5\,GHz is between 13 and 16\,mas and the northern component is seen to
move by 12\,mas over the course of the observations, the observed
movement could in principle have been caused by the self-calibration.
We believe however that the systematic increase in declination of the
northern component seen over a number of days at both frequencies is
indicative of real proper motion.

\subsubsection{Core position}
\label{sec:core}
\citet{Mio01} showed convincing evidence that the northern component
they observed during the 1997 February flare was a core, rather than a
receding counterjet to the extension seen to the south.  The position
of this core in their images appears to be
20$^{\rm{h}}$32$^{\rm{m}}$25$^{\rm{s}}$.7733 +40\degree
57$^{\prime}$27$^{\prime\prime}$.965 (J\,2000).  Comparing this with
the positions in our images shows it to be almost coincident with the
maximum in the northern component of the 22-GHz image from 2001
September 20.  This would seem to argue that this component in our
image (labelled ``C1'' in Figure\,\ref{fig:precession}) is the core.
However, \citeauthor{Mio01} stated that their phase referencing was
not entirely successful, so there is some small uncertainty in their
derived position (although their images prior to self-calibration
showed that the northern component was stationary to within 3\,mas).
Moreover, the 1997 February outburst was 4.5 years earlier than that
of 2001 September.  If the system had a systematic velocity (owing to,
for instance, a natal kick) of 100\,km\,s$^{-1}$, it would move a
distance $\approx95$\,AU in 4.5 years.  If this motion were in the
plane of the sky, that would imply a shift in position of 10\,mas at
the distance of Cygnus X-3.  That this should be along the jet axis is
unlikely, but it should be borne in mind that the source core position
might in principle have changed in the intervening time between the
outbursts.

\subsubsection{Jet precession and modelling}\label{sec:morphology}
\begin{figure}
\psfig{figure=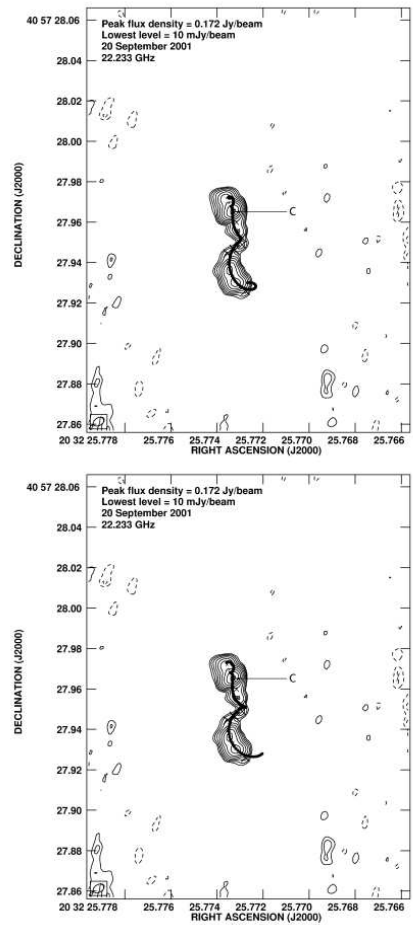,width=8.8cm,angle=0,clip=t}
\caption{Results of fitting the model of \citet{Hje81} to the 22-GHz
  VLBA image of September 20.  The black dots overlaid on the VLBA
  image correspond to the derived positions of knots ejected every 0.1
  days from the core.  The bottom image is for clockwise rotation
  ($s_{\rm rot} = -1$) and the top image for anticlockwise rotation
  ($s_{\rm rot} = +1$).  Both models fit all the major peaks in the
  southern jet, the direction of the northern jet, and the approximate
  lengths of the two jets.  They make no allowance for the fading of
  jet knots with time.
  \label{fig:precession}}
\end{figure}

The final 22-GHz image (from September 20) shows a series of emitting
knots, and the brightest northern component is coincident with the
position of the core as determined by \citet{Mio01}.  We attempted to
fit the precessing jet model of \citet{Hje81} to this image,
investigating the three most plausible core positions, corresponding
to the brightest point in the northern component (at
20$^{\rm{h}}$32$^{\rm{m}}$25$^{\rm{s}}$.77335 +40\degree
57$^{\prime}$27$^{\prime\prime}$.9650), the flux density peak
immediately south of that (at
20$^{\rm{h}}$32$^{\rm{m}}$25$^{\rm{s}}$.77315 +40\degree
57$^{\prime}$27$^{\prime\prime}$.9555), and the flux density peak at
the centre of the jet flow (at
20$^{\rm{h}}$32$^{\rm{m}}$25$^{\rm{s}}$.77295 +40\degree
57$^{\prime}$27$^{\prime\prime}$.9487), all given in J\,2000
co-ordinates.  These positions correspond to the points labelled
``C1'', ``C2'', and ``C3'' respectively in
Figure\,\ref{fig:precession}.  We used a minimisation routine based on
the downhill simplex method \citep{Pre92} in order to find the best
agreement between the model and the image.  The precessing jet model
has eight independent parameters: the inclination of the jet axis to
the observer, $i$; the opening angle of the precession cone,
$\psi$; the angle through which the jet is rotated in the plane of the
sky from an E-W alignment, $\chi$ (related to the position angle by
P.A. $= \chi+90$\degree); the source distance, $d$; the jet speed
$\beta c$; the precession period, $P$; the sense of precession $s_{\rm
rot}$ (clockwise or anticlockwise); and some initial phase of
precession, $\phi$.  We added in an extra parameter $t_{\rm view}$,
the time after ejection at which the image was made.  We measured the
positions and flux densities at fifteen different points along the
spine of the jet, and then weighted the points according to the
corresponding flux densities.  It was found that a simple minimisation
of the sum of the weighted distances between each measured jet
position and the closest model point did not give reasonable
solutions, since the minimised model parameters would predict radio
emission from points well outside the observed jet.  To improve the
fits, additional terms were used in the function to be minimised,
penalising a fit if it predicted too many precession periods to have
passed in the time between ejection and observation, and penalising it
if it predicted radio emission outside a small box containing the
observed jet.  The minimisation routine was run for almost a hundred
thousand different sets of starting parameters, which were chosen in
order to fully sample (at equal intervals) the available parameter
space without taking a prohibitively long time to run.  Results
predicting $\beta>1$, distances outside the range $5<d<15$\,kpc
\citep{Pre00,Dic83}, or negative periods were disallowed.  The sum of
the distances from each jet position to the closest model point was
computed for each of the converged results, and the results with the
lowest such cumulative distances which also predicted a viewing time
$t_{\rm view}<8.5$\,days (the time between the detection of the
initial rise of the radio flux density and the epoch at which the
image was made) were then examined individually, to check that the
results with the lowest cumulative distances did in fact produce good
by-eye fits to the data.  Upon examination, no satisfactory fits were
found for core positions corresponding either to the centre of the jet
flow (core position C3), or to the southern part of the northern
component (C2).  However, good by-eye fits to the data were found for
a core position corresponding to the brightest point in the north
component (C1).  We therefore identify the brightest point in the
northern component as the core of the system.

For that northernmost core position, more good fits were found for
clockwise rotation ($s_{\rm rot}=-1$) than for anticlockwise rotation
($s_{\rm rot} = +1$), and the mean, standard deviation, and the
maximum and minimum values for the satisfactory fits are given in
Table\,\ref{tab:jet_fitting}.  The sense of the jet precession can not
be uniquely determined in this case, but the parameters $\psi$,
$\beta$, $P$, and $t_{\rm view}$ are similar for the two possible
senses.  $\phi$ and $d$ are not uniquely determined, with $\phi$
having a large standard deviation and the distance varying between the
limits allowed by the model-fitting code.  Although the parameters
$i$ and $\chi$ appear very different in the two cases, we believe
that this is due to the way the angles were defined in the model of
\citet{Hje81}.  In changing the sense of jet precession from clockwise
to anticlockwise, all angles will be measured in the opposite sense,
which will affect the definitions of inclination angle and position
angle on the sky.  Redefining $i_{\rm new} = 180-i$ and
$\chi_{\rm new} = \chi - 180$ for the anticlockwise case gives
$i_{\rm new} = 10.4\pm3.8$ and $\chi_{\rm new} = 98.1\pm2.8$, in
good agreement with the parameters derived for the clockwise case.
So, taking the core position to be at
20$^{\rm{h}}$32$^{\rm{m}}$25$^{\rm{s}}$.77335 +40\degree
57$^{\prime}$27$^{\prime\prime}$.9650, as allowed by the fits, we can
constrain the system parameters to be $\psi = 2.4\pm1.2$\degree,
$i = 10.5\pm4.2$\degree, $\chi = 98.9\pm2.9$\degree, $\beta =
0.63\pm0.09$, and $P = 5.34\pm0.67$\,days.  This gives a value
$\beta\cos i = 0.62\pm0.11$.  Comparing this value with that of
$\beta\cos i = 0.50\pm0.10$ derived from the proper motion
analysis of \S\,\ref{sec:proper_motion}, we find a slight discrepancy,
although the error bars do overlap.  Our best model fits for both
senses of precession are shown in Figure\,\ref{fig:precession}.  We
cannot robustly determine the sense of precession with this data.

Thus we do see a two-sided jet on VLBA scales.  However, the jet speed
$\beta$ and the inclination angle $i$ are such that the observed
length of the northern jet is significantly shorter than that of the
southern jet.  The ratio of the arm lengths is
\begin{equation}
\frac{l_{\rm{app}}}{l_{\rm{rec}}} =
\frac{1+\beta\cos i}{1-\beta\cos i},
\end{equation}
where $l_{\rm{app}}$ is the length of the approaching jet and
$l_{\rm{rec}}$ is the length of the receding jet, $\beta$ is the
jet-knot advance speed, and $i$ the inclination of the jet axis
to the line of sight.  The fitted parameters in the precession model
imply a length ratio of $4.3\pm0.3$.  Figure\,\ref{fig:precession}
shows that the north jet is indeed shortened by a factor appropriate
to these values of $\beta$ and $i$, convincingly verifying the
two-sidedness of the jet in \cyg\ on this occasion.

\subsubsection{VLBA component spectra}
\label{sec:vlba_spectra}
The flux densities of the source components (shown in the images of
Figure\,\ref{fig:vlba_images}) during the flare of 2001 September were
measured using the \textit{AIPS} task JMFIT.  The 5-GHz VLBA component
lightcurves are shown in Figure\,\ref{fig:vlba_fluxes}.  The total
emission (the sum of the northern and southern components) was found
to track the VLA flux density.  At 5\,GHz, the integrated flux density
of the northern component exceeded that of the southern component on
September 18 and 19, after which its flux density diminished more
rapidly, and the southern component dominated.  At 22\,GHz, the
northern and southern components could not be distinguished on
September 18.  On September 19, the northern component dominated, and
the flux densities of the two components were comparable on September
20.  When the spectral indices of the VLBA components were determined,
it was found that neither component had an flat or inverted spectrum
between 5\,GHz and 22\,GHz.  Figure\,\ref{fig:precession} shows that
the core and the northern (receding) jet are very close together owing
to relativistic aberration, and therefore would not be resolved at
5\,GHz.  It is the jet which dominates the flux density at low
frequencies (5\,GHz and 15\,GHz), but the core dominates at 22\,GHz
because it has a significantly less steep spectrum than the jet
material.  Therefore spectral indices between these frequencies will
not (without significantly higher resolution) reveal the identity of
the northern emission, be it core or jet.

\subsection{Is the ejection of the VLBA emitting regions coincident
  with the start of the VLA flare?}
\label{sec:flare_start}
The time of the beginning of the radio outburst can be constrained to
being on or prior to JD\,2452167 (2001 September 14), from the
RATAN-600 data \citep{Tru02}, and from Ryle Telescope monitoring
\citep{Poo01}\footnote{http://www.mrao.cam.ac.uk/$\sim$guy}.  The Ryle
Telescope monitoring shows that the 15-GHz flux density began to rise
out of the pre-outburst ``quench'' period from 0.02\,Jy on
JD\,2452162.5 to 0.1\,Jy by JD\,2452165.

Extrapolating the positions of the VLBA knots at 15\,GHz and 5\,GHz
(the absolute 22-GHz positions being inaccurate due to the lack of
phase-referencing) in Figure\,\ref{fig:vlba_positions} backwards in
time, we believe that those knots cannot have been ejected at the time
when the low-resolution telescopes (the RATAN-600 and the VLA)
detected the radio flux density beginning to rise.  Extrapolating to a
precise time of ejection is difficult, owing to the large
uncertainties on the measurements, but it seems that the northern and
southern components would have been coincident perhaps $2.5 \pm 0.5$
days after the RATAN-600 saw the radio flux density start to rise at
the beginning of the outburst.  This is $1.3 \pm 0.5$ days prior to
the first VLBA observation on 2001 September 18.  By way of
explanation, we propose a scenario in which the core flared and only
expelled the jets later.  This we explore in a forthcoming paper.
Note that this is the opposite order of events to that proposed by
\citet{Ato97}, who suggested a scenario whereby the radio flux density
flared well after the ejection of radio clouds, once they had expanded
sufficiently to become transparent at radio frequencies.
Alternatively, it may be that the jet speed increased over the course
of the first few days, although we note that once VLBA monitoring
began (on September 18) the positions of the knots are consistent with
a constant separation velocity, suggesting that their motion has
always been ballistic.

\subsection{Comparison with previous work}

The most detailed previous VLBI images of Cygnus X-3 were made by
\citet{Mio01}.  They made three observations at 15.365\,GHz with the
VLBA at epochs 2, 4, and 7 days after the peak of a 10\,Jy radio flare
in 1997 February.  \citeauthor{Mio01} detected what they interpreted
as a stationary (to within 3\,mas) core component and a continuous jet
extending to the south, reaching a maximum observable extension of
120\,mas four days after the flare, which faded beyond the detection
limits by the last observation.  They took the time of ejection of the
radio-emitting plasma to be the start of the outburst as detected by
the Ryle Telescope and the Green Bank Interferometer (GBI), and from
the measured lengths of the jet at different epochs, derived proper
motions, which constrained the jet speed to be
$\beta_{\rm{min}}\gtrsim (0.78-0.75)$, and a weak upper limit to the
inclination to the line of sight to be $i<78$\,\degree.  Then, from
the ratio of the integrated flux density in the southern jet to the
integrated noise in the north taken over a smaller area (to account
for slower apparent motion of the receding jet), they derived a
constraint on $\beta\cos i\gtrsim0.806$.  We believe that there is
some scope for error in their proper motion analysis, since it rested
on the length of the jet, rather than on tracking individual
well-identified knots of radio emission.  The length of a jet that may
be imaged will be affected by the map quality, which was poor during
at least the first epoch of their observations, owing to bad weather
in the southwestern USA.  Moreover, it is not clear that the radio
plasma was ejected at the start of the Ryle/GBI flare (see
\S\ref{sec:flare_start}).  Uncertainty in the ejection time would then
lead to an error in the proper motion and the derived jet speeds.

A comparison may also be made between the precession parameters
derived for the 1997 flare and those we derived in
\S\,\ref{sec:morphology}.  Their best fits yielded a precession period
$\gtrsim60$ days, a jet speed $\beta\gtrsim0.81$, a precession opening
angle of $\lesssim$12\degree, an inclination angle of
$\lesssim$14\degree\ to the line of sight, and anticlockwise
precession.  The inclination angle, jet speed and opening angle
derived by \citet{Mio01} were similar to those we derived.

Because the derived precession period of the 2001 event is comparable
to the monitoring timescale (c.f.\ the 1997 event, when the monitoring
sampled less than one tenth of the best-fit precession period), we
have confidence in the limits we have placed on this parameter.
However, the fits of \citeauthor{Mio01} robustly require a long
precession period ($\gtrsim 30$ days), whereas ours is of order $\sim
5.3$ days.  It is thus hard to avoid the conclusion that the
precession period has changed by a significant factor between these
two events.

From the above, we believe that the jet speed we derive from proper
motion measurements is consistent with that estimated by
\citet{Mio01}.  In both sets of observations, the southern jet appears
to be brighter and faster-moving than any northern jet and in both
sets of data there is evidence for precession (albeit with rather
different precession parameters).  In addition, the nature of the jet
itself appears to be different, in the sense that the 2001 September
jet seemed to be composed of a few identifiable knots, whereas the
1997 February jet appeared more like a continuous stream of radio
plasma.  On the other hand, intrinsic differences could, in principle,
account for the difference in the observed nature of the jet.  The
outburst of 1997 February was shorter in duration than that of 2001
September.  The integrated flux density died away more quickly in the
1997 flare (falling to below 1\,Jy at 15\,GHz $\sim3$ days after the
peak of the flare, as compared to $\sim5$ days afterwards in 2001),
after which it was still possible to make VLBA observations.  The
dynamic range-limited images of \citeauthor{Mio01} therefore have much
lower r.m.s.\ noise levels than those in
Figure\,\ref{fig:vlba_images}, so lower-level continuous emission (as
opposed to the discrete knots we observed) could be detected.  This
could account for our non-detection of continuous emission, although
the comparison of VLA and VLBA flux densities suggest that we are not
missing significant components in our VLBA images, at least at low
frequencies.

A comparison may also be made to the flare of 2000 September 15
observed by \citet{Mar01}.  They observed \cyg\ with the VLA in its
most extended ``A'' configuration 36, 51, and 66 days after the start
of a radio outburst detected with the GBI.  By this latest time, the
source had faded back to quiescence.  They imaged a two-sided ejection
on arcsecond scales, with a dominant core, a prominent northern
ejection and a strong hint of extension to the south.  The flare
observed with the GBI was less powerful than either the 1997 or the
2001 flare, peaking at only $\sim 7$\,Jy at 8.3\,GHz.

\citet{Mar01} derived proper motions by assuming the ejection date to
be the start of the flare observed with the GBI, and obtained a value
of $\beta\cos i=0.14\pm0.03$.  In fact, their plot of the GBI data
shows several flares of amplitude $>4$\,Jy over the 20 days following
their assumed ejection date (after which no data is plotted, since GBI
monitoring of \cyg\ appeared to cease on JD\,2451821).  This opens up
two possibilities.  It is not clear which (if any) of the plotted
flares was associated with the actual ejection event corresponding to
the jets they observed.  If the last one (JD\,2451822) was the flare
responsible, their third image showing the arcsecond-scale jets would
have been taken only 46 days after the flare.  This would lead to
increased values for the proper motion of the two jet components, and
a derived jet speed of $\beta=0.68$.  It is also possible that the
ejection occurred after their GBI data stops, in which case the
derived proper motions and speeds would be even greater.  A speed of
$\beta=0.68$ is much more consistent with that derived in this paper.
It is not clear however that the expansion rate would be uniform and
it is possible that the jet would slow down as it impinged on its
environment.  While this effect might be negligible over a few days,
it might be significant on a timescale of months.  However, their
value of $\beta\cos i=0.12\pm0.01$ based on the flux density ratio
of the 2 jets interpolated to equal angular distances from the core
would seem to be more robust than the value derived from their proper
motion analysis.

\citeauthor{Mar01} explained the one-sided VLBA jet observed by
\citet{Mio01} by invoking jet obscuration by an absorbing medium on
VLBA scales, in the form of a free-free absorbing disk-like wind.  They
suggested that the northern jet was the approaching jet, and the
southern jet was receding from us.  For the 15-GHz images of
\citeauthor{Mio01}, they derived an optical depth at 50-100\,mas from
the core of $\tau_{\nu}^{\rm{ff}}=1.15\pm0.14$.  Since their quoted
free-free absorption coefficient scales as $\nu^{-2.08}$, that would
imply an optical depth of $0.53\pm0.08$ at 22\,GHz.  This would be
insufficient to prevent us from seeing a bright approaching northern
jet in our images.  Since our observations strongly suggest a receding
northern jet, it would be hard to reconcile their obscuration
explanation with our data.

We made further observations of \cyg\ 134 days after the detected
start of the 2001 September flare, on 2002 January 25, whose detailed
analysis we will present in a forthcoming paper.  As with the
observations of \citeauthor{Mar01}, the VLA was in ``A''
configuration, and the source had faded back to a quiescent flux
density of $\approx100$\,mJy at 43\,GHz, although it displayed
significant variability on a timescale of $\approx 10$\,minutes.  We
subtracted out the time variability, and then imaged the core.
Fitting that with a Gaussian and subtracting this from the
\textit{uv}-data set, there was no evidence of any extension similar
to that seen by \citeauthor{Mar01}, to an r.m.s.\ limit of 0.8\,mJy
per beam.  We note, however, that this is as expected, since
\cite{Mar01} detected northern and southern extensions 66 days after
the assumed ejection event (see discussion above) at levels of
1.45\,mJy and 0.89\,mJy respectively.  Assuming that the flux
densities of these extensions decreased with time, then 134 days after
the start of the outburst, such extensions would be undetectable at
our r.m.s.\ limit.

\section{Radio spectra}\label{sec:spectra}
\subsection{General characteristics}

\begin{figure}
\psfig{figure=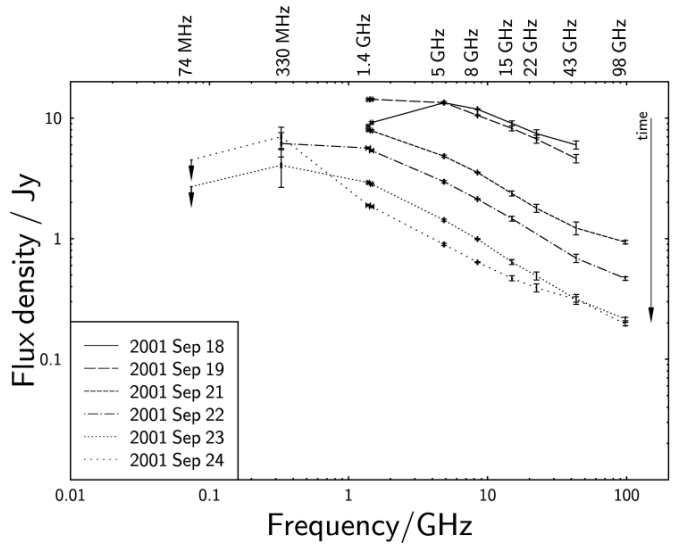,width=8.8cm,angle=0,clip=t}
\caption{Evolution of the radio spectrum of \cyg\ \label{fig:spectra}}
\end{figure}

\begin{figure}
\psfig{figure=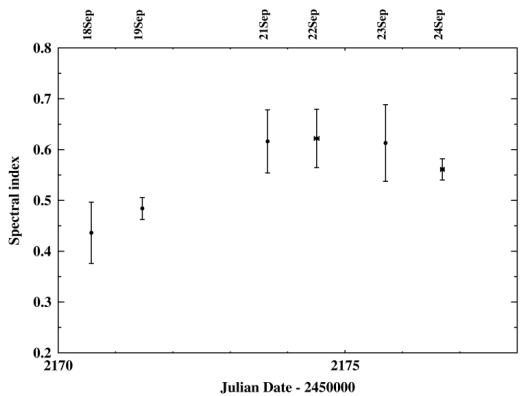,width=8.8cm,angle=0,clip=t}
\caption{Evolution of the spectral index of \cyg.  The spectral index
  was fitted with a simple power-law for the portion of the spectrum
  that was well above the low-frequency turnover ($>8.4$\,GHz for
  September 18 and 19, $>5$\,GHz for September 21 and 22, and
  $>1.4$\,GHz for September 23 and 24).
  \label{fig:specindex}}
\end{figure}

The high-frequency parts of the spectra shown in
Figure\,\ref{fig:spectra} were fitted with single power laws and the
resultant spectral indices are shown in Figure\,\ref{fig:specindex}.
The spectral index evolved during the course of the flare, from a
value of $0.44\pm0.06$ on September 18 to a value of $0.61\pm0.08$ by
September 23.  Traditional models of synchrotron radiation cannot
explain such a spectral steepening, although this phenomenon has been
seen before, in the giant radio outburst of 1972 \citep{All72, Hje72,
Pet73}, and in subsequent outbursts \citep[e.g.][]{Sea74, Gel83}.

The spectra also show that by September 24 (if not earlier), the
spectra began to deviate from pure power laws, and rapid variability
began to affect the source flux.  Two different scans were made at 8.4
GHz on September 24, each of duration 2 minutes, and separated by 30
minutes (this was the only epoch when separated scans were made); in
this time, the source flux density increased from $618.2\pm1.2$\,mJy
to $656.5\pm1.2$\,mJy.  The main outburst dominated the source flux
during the first few days of observation, but as the outburst decayed,
the source behaviour was affected more by core variability or minor
ejections.

\subsection{What causes the low-frequency turnover?}
The low-frequency (1.4-GHz to 74-MHz) measurements indicate a turnover
in the radio spectrum.  We now consider various possible mechanisms
for creating such a turnover, and assess the plausibility of each one.
Such low-frequency turnovers in other sources are generally ascribed
to one of four distinct mechanisms \citep{Kel66, Gre74}.  These are:
synchrotron self-absorption (SSA), a low-energy cutoff in the electron
energy distribution spectrum, the Tsytovich-Razin (T-R) effect, and
free-free absorption (FFA) by thermal plasma, either with the plasma
uniformly mixed with the radiating electrons or as intervening
material along the line of sight to the source.  We now discuss these
four possibilities in turn.

\subsubsection{Synchrotron self-absorption} \label{sec:ssa}

Well below the turnover frequency $\nu_{\rm m}$, where the optical
depth $\tau \gg 1$, a synchrotron self-absorbed spectrum scales as
$S_{\nu} \propto \nu^{5/2}$ \citep[e.g.][]{Ryb79}.  In the
optically-thin region however, where $\tau\ll 1$, the scaling is
$S_{\nu} \propto \nu^{-\alpha}$ ($\alpha>0$).  The width of the
turnover region depends on the structure of the source.
Unfortunately, we do not sample the low-frequency spectrum well enough
to determine whether the terminal low-frequency spectrum scales as
$\nu^{5/2}$.  However, we clearly observed that the turnover moved to
lower frequencies with time.  If the synchrotron self-absorption
paradigm is correct, this would be consistent with the observed
expansion of the plasmons, which would cause the magnetic field within
the plasmons to decline, and thus lower the turnover frequency.
Magnetic flux freezing implies that $B/\rho\propto\Delta l$
\citep{Lan63}, where $\rho$ is the plasma mass density, $B$ is the
magnetic flux density, and $\Delta l$ is the length of a fluid element
along the magnetic field line.  Assuming that the mass of the plasma
cloud is constant and that there is no strong turbulence, $\Delta
l(t)\propto R(t)$, and $B\propto R^{-2}$.  However, if turbulence
develops in the plasma, the fluid elements are stretched by turbulent
eddies in addition to the expansion of the cloud, so $\Delta l\propto
R^{1+\alpha}$, where $\alpha$ in this case is a factor describing the
degree of turbulence, typically taken as of order 1 for a turbulent
plasmon with a correspondingly tangled magnetic field.  In this case,
$B\propto R^{-2+\alpha} \approx R^{-1}$.  So as the plasmon expands,
the magnetic field declines, and the turnover frequency decreases.
Thus the SSA scenario is consistent with our data.

\subsubsection{Low-energy cutoff in the electron spectrum}
We modelled the electron distribution as a power law.  If that power
law were to extend only down to some minimum energy, below which the
electron distribution turned over, producing a cutoff in the electron
distribution function, that would produce a low-frequency turnover in
the radio spectrum.  This turnover would move to lower frequencies as
the energies of the individual electrons decreased due to adiabatic
expansion of the plasmon, moving the cutoff in electron energies to
lower energy.  We cannot rule out this possibility with our data.
Depending on the abruptness of the cutoff, observing a $\nu^{1/3}$
spectrum below the turnover would identify this scenario.

\subsubsection{The Tsytovich-Razin effect}
The refractive index, $n$, of a plasma containing free electrons and
ions may differ from unity and signals then travel at the phase
velocity $c/n$.  In a medium where the refractive index is less than
one, beaming from relativistic electrons is suppressed, as the angle
into which the radiation is beamed is given by
\begin{equation}
1/\gamma = \sqrt{1-n^2\beta^2}.
\end{equation}
At low frequencies ($\nu \ll \nu_{\rm{p}}$, where $\nu_{\rm{p}}$ is the plasma
frequency), this effect is dominant, and the synchrotron spectrum is
cut off due to beaming suppression.  At higher frequencies, this
effect is unimportant, since
\begin{equation}
n^2=1-\left(\frac{\nu_{\rm{p}}}{\nu}\right)^2.
\end{equation}
The frequency in GHz at which this effect becomes significant is
given by
\begin{equation}
\nu \sim 2\times10^{-14}n_{\rm{th}}/B,
\end{equation}
where $B$ is the magnetic field in Gauss and $n_{\rm{th}}$ is the density
of ambient thermal matter in m$^{-3}$.  In \S\,\ref{sec:mag_field} we
estimate an upper limit on the magnetic field to be $B \sim 0.15$\,G;
thus if the T-R effect were the mechanism by which the
spectrum turns over, a rather high (upper limit to the) density of
$8 \times 10^{12} {\rm m}^{-3}$ of thermal plasma would be required
for a turnover frequency of 1\,GHz.  At these densities, free-free
absorption would also be expected to become important.

\subsubsection{Free-free absorption}
We consider two possible geometries for free-free absorption.  The
absorbing thermal material could either form a screen enclosing but
not permeating the source, or it could be mixed in with the
synchrotron-emitting plasma.

In the former case, if the plasmons were expanding into a region of
thermal material, then the flux of the receding component would fall
off more rapidly with time than the flux of the approaching component,
since the length of the absorbing screen would increase as the
receding plasmon moved away from the core.  The length of the
absorbing screen between the observer and the approaching component
would decrease with time, so the turnover would move to lower
frequencies with time as this component began to dominate the observed
flux density.  In fact, we see evidence that the 5-GHz flux density of
the northern (jet) component falls off rapidly with time after
September 19, whereas that of the southern component does not diminish
so rapidly.  This is then in principle a plausible scenario.

For thermal material to be mixed with the synchrotron-emitting plasma,
the material would either need to be present at the time the plasmon
was ejected, or would have to be entrained during the subsequent
motion of the plasmon through the surrounding medium.  In order to
create a turnover at frequencies as high as $\sim 3$\,GHz by September
18, the initial rate of entrainment would have to be extremely high.
To cause the observed decrease in turnover frequency, the rate would
then have to be quite accurately tuned.  We do not favour this
explanation.

\subsubsection{Discussion on the origin of the turnover}
The change in turnover frequency with time places more stringent
constraints on the mechanism responsible for the turnover.  For the
T-R effect, the change in turnover frequency with time would require a
change in $n_{\rm{th}}/B$.  For FFA, it requires a decreasing density
with time, either due to the source moving into lower-density regions
as it moves outwards, or due to the expansion of thermal material
mixed in with the synchrotron-emitting electrons.  For SSA, expansion
and a corresponding decrease in the magnetic field lead to a decrease
in the absorption coefficient and therefore in the turnover frequency.
Multi-frequency imaging with sufficiently high resolution at or below
the turnover frequency could in principle distinguish definitively
between the mechanisms.

\subsection{High-frequency spectral evolution}
In order to explain the high-frequency spectral evolution, in
\S\,\ref{sec:analysis}, we develop a model which factors in light
travel time effects both within and between components, which we
believe explains the basic characteristics of the spectral evolution,
e.g.\ the observed timescale.  But first we derive some basic
parameters of the source in \S\,\ref{sec:src_parms}.

\section{Constraining the magnetic field and Lorentz factors}\label{sec:src_parms}
\subsection{Magnetic Field}\label{sec:mag_field}

If self-absorption were the mechanism responsible for the turnover,
precise size constraints from VLBA measurements would allow the direct
determination of the magnetic field.  If, however, another process
were responsible, the self-absorption turnover frequency $\nu_{\rm m}$
would be at a lower frequency than the observed turnover in the
spectrum.  Thus an extrapolation of the power-law spectrum back to the
predicted self-absorption turnover frequency would give a higher
extrapolated turnover flux density $S_{\rm m}$ than is observed.
Since $B \propto S_{\rm m}^{-2}\nu_{\rm{m}}^5$, assuming that the
observed turnover frequency and flux density are due to SSA still
gives an upper limit to the magnetic field strength.

Assuming that the turnover in the spectrum of \cyg\ is due to
SSA, then given the frequency, $\nu_{\rm{m}}$,
and flux density, $S_{\rm{m}}$, of the turnover, and the angular size
$\theta$ of the emitting region, synchrotron theory gives an
expression for the magnetic field in that region \citep{Mar83}.
\begin{equation}
B=10^{-5}b(\alpha) \theta^4 \nu_{\rm{m}}^5
S_{\rm{m}}^{-2}\left(\frac{\delta}{1+z}\right)\textrm{G}, \label{eq:bfield}
\end{equation}
where $S_{\rm{m}}$ is measured in Jy, $\theta$ in milliarcseconds and
$\nu_{\rm{m}}$ in GHz.  The parameter $b(\alpha)$ is a slowly-varying
function of the high-frequency spectral index with a value of
$\approx 3.4$ for $\alpha=0.6$, and $\delta$ is the beaming parameter
\begin{equation}
\delta = \frac{1}{\Gamma(1-\beta\cos i)},
\end{equation}
where $\Gamma$ is the bulk Lorentz factor, $(1-\beta^2)^{-1/2}$.  The
redshift $z$ is zero.

Using the VLBA observations to constrain the number and sizes of
synchrotron-emitting components, it is possible to fit spectra to
those components and derive upper limits to their magnetic fields.
Unfortunately, the strong dependence of the magnetic field on the
angular diameter $\theta$ and the turnover frequency $\nu_{\rm{m}}$
means that the uncertainty in the derived field is quite high.  It
also makes it extremely important to define correctly the angular
diameter $\theta$.  The radiative transfer equation used to derive
Equation \ref{eq:bfield} assumes spherical components.  The fitting
task in \textit{AIPS} used to derive the sizes models the components
as elliptical Gaussians.  \citet{Mar87} gives an expression for the
value $\theta_{\rm{s}}$ of a sphere which subtends the same solid
angle as an ellipse of major and minor FWHM axes $\theta_{\rm{G_a}}$
and $\theta_{\rm{G_b}}$ as
\begin{equation}
\theta_s = 1.8\sqrt{\theta_{\rm{G_a}}\theta_{\rm{G_b}}}. \label{eq:theta}
\end{equation}
These equivalent angular sizes are listed in
Table\,\ref{tab:vlba_sizes}.  Using Equations \ref{eq:bfield} and
\ref{eq:theta}, together with the fitted turnover frequency of
$3.01\pm0.06$\,GHz, the turnover flux density of $15.1\pm0.5$\,Jy, the
spectral index of 0.4, and the measured core size of
$7.88\pm0.51$\,mas on September 18 (when the VLBA observations
suggest that the source consisted of a single component; we note that
even if it did not, the individual component sizes would then be
smaller than the fitted size, which we could therefore still use to
place a reasonably firm upper limit on the magnetic field), an upper
limit on the magnetic field of $1.5\pm0.6\times 10^{-1}$\,G
($1.5\pm0.6\times 10^{-5}$\,T) is derived.

This is our best estimate of the magnetic field strength.  It could,
however, be at least an order of magnitude lower.  The proper motion
of the plasmons during the course of a VLBA observation would elongate
the observed knots in the images and cause us to overestimate their
sizes.  If this effect were taken into account, the derived magnetic
field might be at least as low as $1.94\pm0.96\times 10^{-2}$\,G.

\subsubsection{Equipartition?} \label{sec:equipartition}
If the particle and magnetic field energies in the source components
were close to equipartition, it would be possible to estimate the
minimum-energy magnetic field in the plasmon.  Following the
derivation set out in \citet{Lon94}, and using measurements of the
flux densities and plasmon sizes (Table\,\ref{tab:vlba_sizes}) from
September 18, and assuming a factor of 100 times more energy in
relativistic protons than in electrons, this assumption gives a
minimum-energy magnetic field of $6\times10^{-2}$\,G, corresponding to
a minimum energy in the plasmon of $3\times10^{34}$\,J (an energy
density of $4\times10^{-5}$\,Jm$^{-3}$).  Since our upper limit on the
magnetic field is greater than the minimum-energy field, our
measurements are not necessarily inconsistent with equipartition
(although we note that there is no compelling reason why the
equipartition assumption should hold in such a manifestly non-steady
state system).

\subsubsection{Comparison to other magnetic field estimates}

Most derivations of the magnetic field in microquasar jets come from
equipartition arguments, despite the uncertainty as to whether
equipartition is valid in these systems.  \citet{Spe86} used
equipartition arguments to put a lower limit of 0.08\,G on the
magnetic field of the emitting components during the 1983 September
outburst of Cygnus X-3.  Equipartition assumptions applied to other
sources have also led to minimum-energy magnetic fields of order
$0.01-1$\,G (for instance GRS\,1915+105 \citep{Ato99}, LS\,5039
\citep{Par00}, and Scorpius X-1 \citep{Fom01}).

A few magnetic field estimates are available in the literature which
do not rely on equipartition.  \citet{Ogl98} used electron ageing
arguments to put an upper limit on the magnetic field in the jet of
Cygnus X-3 in quiescence of $B\leq20$\,G at $5\times10^{12}$\,cm from
the core.  \citet{Par02} considered inverse Compton processes from a
conical jet to constrain the magnetic field at 1000\,AU from the core
of LS\,5039 to lie between 0.06 and 128\,G.  Thus our estimate of the
magnetic field in Cygnus X-3 is consistent with derived values in both
this system and in other microquasars.

\subsection{Lorentz factor}
\label{sec:lorentz}
To a good approximation the radiation observed at a given frequency
$\nu$ probes electrons of Lorentz factor
\begin{equation}
\gamma=\left(\frac{2\pi m_{\rm{e}}\nu}{eB}\right)^{1/2},\label{eq:nu_gamma}
\end{equation}
where $B$ is the magnetic field in T, $\nu$ the observing frequency in
Hz, and the constants are in S.I.\ units.  In the case of adiabatic
expansion, the magnetic field decreases as the source expands, and
observations at a fixed frequency sample electrons with higher Lorentz
factors.

From the derived upper limit to the magnetic field on September 18 of
0.15\,G, and the maximum and minimum observing frequencies on that day
(43.3 and 1.4\,GHz respectively), the lower limits to the Lorentz
factors of the electrons probed by the measurements are 321 and 58
respectively.

\subsection{The $B-\gamma$ plane}
Synchrotron theory, taken together with our data, can be used to
derive constraints on the area of the $B-\gamma$ plane
(Figure\,\ref{fig:b_gamma}) occupied by the source in 2001 September.

\subsubsection{Synchrotron self-absorption turnover} \label{sec:ssat}
The upper limit on the magnetic field derived in
\S\,\ref{sec:mag_field} from synchrotron self-absorption arguments is
shown as line 1 in Figure\,\ref{fig:b_gamma}.

\subsubsection{Energetics} \label{sec:bg_energetics}
If we assume that the energy ejected during the outburst cannot
greatly exceed the energy emitted by a body radiating at the Eddington
luminosity over the rise time of the outburst (taken to be of order 1
day), then an upper limit on the total energy in the plasmon should be
of order $\sim10^{36}(M/M_{\odot})$\,J.  This energy is contained in
the relativistic particles and the magnetic field.  An upper limit to
the magnetic field can be found if all the energy is assumed to be in
the field rather than the particles.  Knowing the volume of the
plasmon (assumed spherical) on September 18, this leads to a limit on
the energy density of $10^{-3}(M/M_{\odot})$\,Jm$^{-3}$.  By measuring
the Doppler shifts of X-ray spectral lines, \citet{Sta03} placed an
upper limit on the mass of the compact object of $3.6$\,\msun (subject
to certain assumptions).  Thus an upper limit of $6\times10^{-1}
(M/M_{\odot})^{1/2} = 1.1\,$\,G may be placed on the magnetic field
(shown as line 2 in Figure\,\ref{fig:b_gamma}).

\subsubsection{Absence of catastrophic synchrotron losses} \label{sec:bg_synch_losses}
The fact that the spectrum shows no noticeable break up to 100\,GHz
implies that synchrotron losses cannot be important up until at least
the time of the last observation on September 24.  The flare began on
September 14.  Thus the synchrotron loss timescale $t_{\rm{syn}}$ must
exceed ten days.  This puts a limit on $B^2\gamma$, namely
\begin{equation}
B^2\gamma<8.9\times10^{2} \quad\rm{G^2}. \label{eq:tsynch}
\end{equation}
This is shown as line 3 in Figure\,\ref{fig:b_gamma}.

\subsubsection{Lack of leakage} \label{sec:bg_leakage}
Another constraint may be derived by considering the
synchrotron-emitting electrons.  To emit synchrotron radiation, they
must be gyrating in a magnetic field.  In order for the electrons to
remain within the plasmon (where the magnetic field is high enough for
them to produce detectable radio emission), their gyroradius must be
significantly smaller than the size of the plasmon.  If it becomes too
large, electrons will begin to leak out of the plasmon and will no
longer produce detectable emission.  The gyroradius $r_{\rm{g}}$ of an
electron of Lorentz factor $\gamma$ in a field $B$ is
\begin{equation}
r_{\rm{g}} = \frac{\gamma m_{\rm{e}}c}{eB}.
\end{equation}
Requiring this to be less than some fraction $f$ of the size of the
plasmon gives
\begin{equation}
\frac{\gamma m_{\rm{e}}c}{eB}<f\theta d,
\end{equation}
where $\theta$ is the observed angular size of the plasmon and $d$ is
the distance to the source (taken as 10\,kpc).  This limit is shown as
line 4 in Figure\,\ref{fig:b_gamma}.

\subsubsection{Region probed by the measurements}
\begin{figure}
\psfig{figure=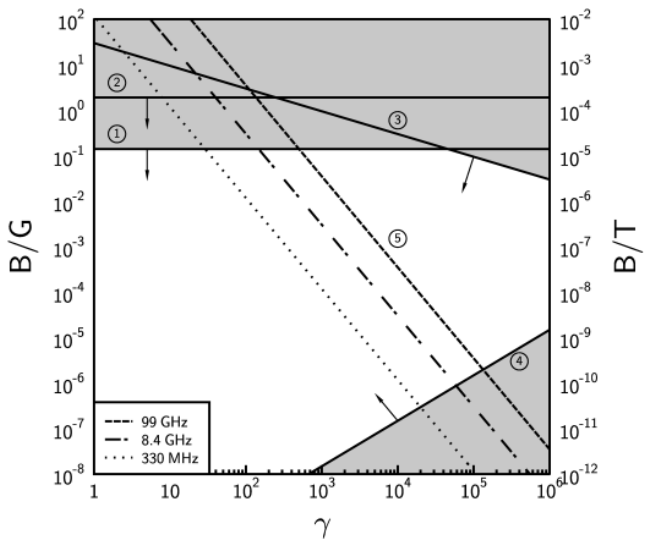,width=8.8cm,angle=0,clip=t}
\caption{Constraints on the $B-\gamma$ plane.  1 is the upper limit on
  the magnetic field from SSA (Equation\,\ref{eq:bfield}) and the size
  constraint of 7.7\,mas from VLBA observations (\S\,\ref{sec:ssat}).
  2 is the constraint from energetics considerations assuming a 3.6
  solar mass compact object (\S\,\ref{sec:bg_energetics}).  3 is the
  synchrotron ageing limit from Equation\,\ref{eq:tsynch}
  (\S\,\ref{sec:bg_synch_losses}).  4 is the limit from the
  requirement for the gyroradius to be less than 10\% of the plasmon
  size, for an angular size of 7\,mas (\S\,\ref{sec:bg_leakage}).  5
  are the loci in the plane probed by observations at three given
  frequencies (Equation\,\ref{eq:nu_gamma}).  Dashed lines are for
  99\,GHz, dot-dashed lines for 8.4\,GHz and dotted lines for
  330\,MHz.  The arrows show which side of the constraint lines are
  valid parameter space.  The shaded areas denote parameter space
  which is excluded.
  \label{fig:b_gamma}}
\end{figure}

Since the electrons radiate at particular frequencies,
Equation\,\ref{eq:nu_gamma} can be used to find the exact lines on the
$B-\gamma$ plane probed by the observations at each frequency (shown
as lines 5 in Figure\,\ref{fig:b_gamma}).

\subsection{Number and energy densities}
\label{sec:num_density}
From \citet{Lon94} and integrating over a power-law distribution of
electrons with exponent $p \neq 2$ (in acknowledgement of the terminal
high frequency spectral indices we observe, even though $p=2$ is the
most frequently-quoted case as it simplifies the mathematics) an
expression for the number density of emitting electrons may be
derived.  Modelling the emitting region as a sphere of uniform
density, of radius $r = \case{1}{2}\theta d$, the number density (in
m$^{-3}$) of emitting electrons is
\begin{equation}
n_{\rm{e}} =
\frac{24S_{\nu}(m_{\rm{e}}c^2)^{1-p}(\gamma_{\rm{min}}^{1-p} -
  \gamma_{\rm{max}}^{1-p})}{2.344\times 10^{-25}(p - 1)d\theta^3a(p)\left(\frac{B}{10^4}\right)^{\frac{p+1}{2}} \left(\frac{1.253\times
    10^{37}}{\nu}\right)^{\frac{p-1}{2}}}\label{eq:number_density}
\end{equation}
where $d$ is the distance to the source in metres, $\theta$ is the
angular size of the source in radians, $S_{\nu}$ is the observed flux
density (measured in Wm$^{-2}$Hz$^{-1}$) of the source at frequency
$\nu$ (Hz), $B$ is the magnetic field of the source in G, $m_{\rm{e}}$
is the mass of the electron in kg, $c$ is the speed of light
(ms$^{-1}$) and $a(p)$ is a slowly-varying function of $p$ which is
tabulated in \citet{Lon94}.  $\gamma_{\rm min}$ and $\gamma_{\rm max}$
are the minimum and maximum Lorentz factors over which the electron
spectrum is known to extend, and are thus the limits of integration
used in calculating Equation\,\ref{eq:number_density}.

From the September 18 data, Equation\,\ref{eq:number_density} gives a
lower limit on the electron density of $9.2\times10^7\,\rm{m}^{-3}$,
although we note that this is uncertain for many reasons not least the
limits of integration used (we used only the observed range of Lorentz
factors listed in \S\ref{sec:lorentz}) and the uncertainty in the
sizes of the plasmons and hence the derived magnetic field.  A lower
limit on the energy density in relativistic electrons therefore is
$9.8\times10^{-4}\,\rm{Jm}^{-3}$.  An upper-limit on the energy
density in the magnetic field is estimated (from the upper limit to
the field derived in \S\ref{sec:mag_field}) to be
$9.0\times10^{-5}\,\rm{Jm}^{-3}$ on September 18.  This gives a ratio
of energy in particles to energy in the magnetic field of $\gtrsim10$
for that day.  The magnetic field would then be a factor 3 less than
predicted by equipartition arguments.

\section{High-frequency spectral evolution}\label{sec:analysis}

The evolution of synchrotron-emitting plasma will be influenced by
adiabatic expansion of the emitting region \citep{vanderLaa66, Sch68,
Hje88}.  The energy loss rate for a gas of relativistic particles in
an expanding volume is
\begin{equation}
\dot{E} = -\case{1}{3}(\nabla\cdot\textbf{v})E,
\end{equation}
where $E$ is the particle energy and $\textbf{v}$ is the particle
velocity.  For a uniformly-expanding sphere of radius $R$, the radial
velocity distribution in the sphere is
\begin{equation}
v=\dot{R}(r/R),
\end{equation}
so that
\begin{equation}
\dot{E}=-\frac{\dot{R}}{R}E.
\end{equation}
The energy loss rate is proportional to the energy of the particle, so
that the expansion is self-similar, and the spectral shape is
therefore preserved.

We observed an initial high-frequency spectrum
$S(\nu)\propto\nu^{-0.43\pm0.06}$ which steepened between September 18
and September 21 to approximately $S(\nu)\propto\nu^{-0.6}$, and then
remained consistent with this terminal value for the remaining three
days of observation.  To cause the spectrum to steepen, some form of
energy-dependent loss mechanism must be invoked.  We now consider
several possibilities for this.

\subsection{Synchrotron losses}
The energy loss rate for a synchrotron-emitting electron is 
\begin{equation}
\dot{E} = -\frac{4}{3}\sigma_{\rm{T}} c U_{\rm{mag}}\gamma^2.
\end{equation}
Therefore the synchrotron loss timescale is given by 
\begin{equation}
t_{\rm{syn}} \equiv -\frac{\gamma}{\dot{\gamma}} \approx
\frac{1}{1.3\times10^{-9}B^2\gamma} \quad\rm{seconds}, \label{eq:t_synch}
\end{equation}
where $B$ is measured in Gauss.  The upper limit on the magnetic field
of order $10^{-1}$\,G (\S\ref{sec:mag_field}) leads to a lower limit
on the synchrotron loss timescale $t_{\rm{syn}}$ of order 11 years.
So synchrotron losses cannot be responsible for the steepening over a
timescale of order days.  Moreover, synchrotron losses are predicted
to create a break in the spectrum which moves to lower frequencies
with time.  We do not observe a break in the high-frequency spectrum,
but rather a smooth spectrum with a single spectral index all the way
up to 100\,GHz, albeit one which initially varies with time.

\subsection{Bremsstrahlung losses}

The bremsstrahlung energy loss rate for an ultrarelativistic electron
in a fully-ionised hydrogen plasma of number density
$n_{\rm{H}}$\,m$^{-3}$ is
\begin{equation}
\dot{E} = -7\times10^{-23}n_{\rm{H}}(\log\gamma+0.36)E
\end{equation}
which leads to an energy-loss timescale of 
\begin{equation}
t_{\rm{brem}} = 6.1\times10^{21}n_{\rm{H}}^{-1} \quad\rm{seconds}
\end{equation}
for particles of Lorentz factor $10^2$.  In order to give a cooling
time of order days, a proton density of $1.6\times
10^{16}\,\rm{m^{-3}}$ is required.  This is nine orders of magnitude
greater than the electron number density inferred in
\S\,\ref{sec:num_density} and would imply a total baryonic mass almost
equal to a solar mass in the plasmon.  This proton density would give
rise to a very large emission measure, and hence the free-free opacity
would be very high, and we would not see the observed synchrotron
emission.  We therefore rule out bremsstrahlung as the loss mechanism
responsible for the spectral evolution.


\subsection{Inverse Compton radiation losses}

An energetic electron in a radiation field of energy density
$U_{\rm{rad}}$ loses energy at the rate
\begin{equation}
\dot{E} = -\case{4}{3}\sigma_{\rm{T}}cU_{\rm{rad}}\gamma^2.
\end{equation}
Hence the cooling time for inverse Compton radiation is
\begin{equation}
t_{\rm{IC}} \equiv -\frac{\gamma}{\dot{\gamma}} \approx
3.1\times10^6U_{\rm{rad}}^{-1}\gamma^{-1}.
\end{equation}
For a particle of Lorentz factor $\gamma\sim10^2$, a cooling time of
order a few days would require a radiation field of order
$U_{\rm{rad}} \approx 10^{-1}$\,Jm$^{-3}$ in soft photons.  Given this
energy density, a source of photons local to the plasmons would be
required, since background radiation fields (the CMB, the interstellar
background etc.) would not have the required intensity.

If the electrons inverse-Compton scatter off an external source of
photons of luminosity $L$ emitted a distance $r$ away, then if the
plasmon is not moving relativistically,
\begin{equation}
U_{\rm{rad}}=\frac{L}{4\pi c r^2},
\end{equation}
where all quantities given are in S.I.\ units.  Inserting the energy
density we require in order to have an inverse Compton cooling time of
order a few days gives a constraint on the photon luminosity we need
and the distance of the plasmon from that source,

\begin{equation}
\frac{L}{L_{\rm{Edd}}}\approx 6.6\times 10^2\left(\frac{r}{\rm
  100\,AU}\right)^2,
\end{equation}
where $L_{\rm Edd}$ is the Eddington luminosity for a ten solar mass
companion star.  Thus, we do not consider this as the likely dominant
loss mechanism, since the companion star's luminosity would have to
exceed its Eddington luminosity by at least two orders of magnitude.

\citet{Fen97} reported that inverse Compton losses were initially
dominant over expansion losses at 15GHz during a flare in 1994
February.  Their model showed that the relative significance of the
inverse Compton losses then declined to $\sim 10\%$ of the expansion
loss level by three days after injection.  We note that they assumed a
significantly slower jet velocity ($\sim 0.3c$) than we found in
\S\,\ref{sec:morphology}.  They also proposed that radiation from the
stellar wind of the companion star would contribute to the radiation
field (a plausible assumption if the companion is indeed a Wolf-Rayet
star).  The combination of these two effects could in principle
account for the lower significance that we attribute to inverse
Compton losses.

\subsection{Leakage}
Energy-dependent escape of relativistic electrons from the plasmons
was considered by \citet{Ato99} in order to explain the spectral
steepening observed in the 1994 outburst of GRS\,1915+105.  They also
put forward a second possibility, that the steepening was caused by
the continuous injection of electrons with a gradually steepening
electron energy spectrum.  Our VLBA observations of \cyg\ seem to
indicate a discrete set of ejecta, so we discount the continuous
ejection scenario on observational grounds.  We now consider the first
possibility, that of energy-dependent escape.

Electrons within the plasmon will be scattered by turbulent magnetic
fields and diffuse through the cloud as a result. This will cause
diffusive leakage from the surface of the plasmon on a timescale $R^2/D$
where $R$ is the plasmon size and $D$ the energetic electron diffusion
coefficient which is $D=\lambda c/3$ where $\lambda$ is the mean free
path. In quasilinear theory the smallest mean free path allowed is the
particle gyroradius, which corresponds to a turbulent field of order
the mean field,

\begin{equation}
r_{\rm{g}}=\frac{\gamma mc}{eB}=1.7\times 10^{-3}\frac{\gamma}{B},
\end{equation}
and hence
\begin{equation}
\left(\frac{r_{\rm{g}}}{77\,{\rm{\,AU}}}\right)=1.4\times 10^{-11}
\gamma\left(\frac{B}{10^{-1}\rm{\,G}}\right)^{-1},
\label{eq:gyro}
\end{equation}
which gives, roughly, a gyroradius (in units of the plasmon size) of $10^{-9}$.
The diffusive escape time with such a mean free path is then 
\begin{equation}
\begin{split}
t_{\rm{esc}}&=\frac{R^2}{D(\gamma)}\\
\left(\frac{t_{\rm{esc}}}{10\,\rm{years}}\right)&\approx
\left(\frac{\gamma}{10^2}\right)^{-1}
\left(\frac{R}{77\,\rm{AU}}\right)^2
\left(\frac{B}{10^{-1}G}\right).
\end{split}
\end{equation}

Spectral steepening cannot be due to diffusive escape if the
scattering occurs on the scale of a gyroradius.  A leakage timescale
of a few days would require a mean free path of
\begin{equation}
\lambda \sim 10^{-1}R,
\end{equation}
thus leakage via this mechanism is not a plausible explanation for the
observed energy losses, given the gyroradius derived in
Equation\,\ref{eq:gyro}.

\subsection{Light travel-time effects}
\label{sec:ltt}
We now consider an alternative explanation for the observed
high-frequency spectral evolution, namely that it is governed by light
travel-time effects in a series of plasmons which are ejected and
radiate at different distances from the observer, and evolve from
being optically-thick to optically-thin as they move outwards from the
core and expand (in fact the effect will be important whether or not
the plasmons expand, as long as they have a finite size and get fainter
with time.  We invoke expansion, since that is implied by our
observations, as discussed in \S\,\ref{sec:plasmon_size}.)  At 5\,GHz
(the frequency at which we have best temporal coverage of VLBA
images), the separation of the northern and southern emitting regions
expands from a projected size of 12\,mas on September 18 to 39\,mas on
September 21.  For a source distance of 10\,kpc, this equates to
120\,AU on September 18 and 390\,AU on September 21.  The sizes of the
individual plasmons increase from 7\,mas to 17\,mas, equating to 70 and
170\,AU respectively.  Thus it takes light a minimum of between 9 and
22\,hours respectively to travel across each plasmon, and between 16
and 52\,hours to travel between the plasmons.  The source is evolving
on a similar timescale, so light travel-time effects \textit{must} be
relevant.  Since the inclination of the jet axis to the line of
sight has been found to be small ($i=10.5\pm4.2$\degree,
\S\,\ref{sec:morphology}), light travel-time effects between the
plasmons will be important.  At any one time, we see light from the
far side of a given plasmon at an earlier epoch than light emitted
from its near side.  At earlier epochs the plasmon's magnetic field
and emissivity were higher.  As the plasmon expands, the field,
emissivity, and any opacity decrease.  If the combination of size and
magnetic field are such that opacity effects are initially important,
then initially an overall flat (or inverted) spectrum would be seen.
During the transition from $\tau \gg 1$ to $\tau \ll 1$, the near side
of a plasmon will contribute optically thin emission, but we will only
see in to an optical depth of $\tau\approx1$.  After sufficient time
has elapsed, when all of the plasmons are optically thin, a constant,
steep, terminal value of the spectral index will be observed.

\begin{figure}
\psfig{figure=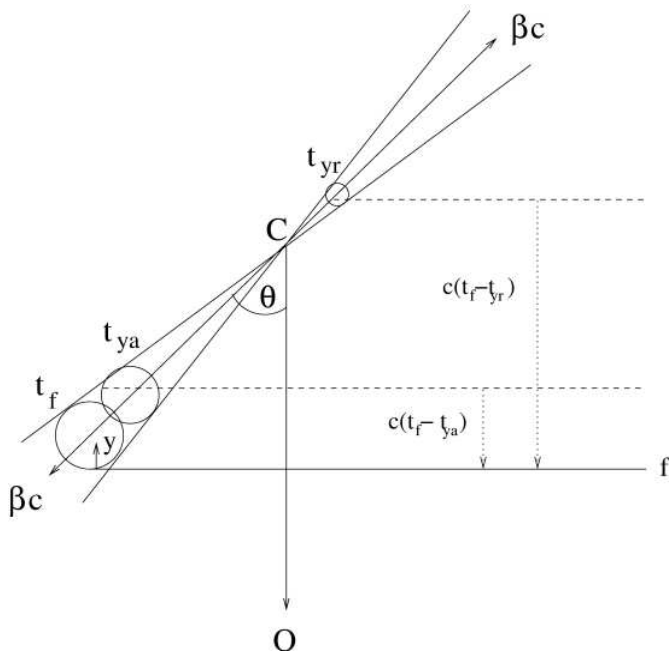,width=8.8cm,angle=0,clip=t}
\caption{Assumed geometry of the jet. C is the core of the system, and
  the observer is in direction O.  The plasmon centres move outwards
  at speed $\beta c$ and the jet axis is aligned at angle $i$ to
  the line of sight.  We use the co-ordinate $y$ as a dimensionless
  local measure of the distance from the front surface to any plane
  within a given plasmon.  Light from position $y$ within a plasmon
  leaves that plasmon in the approaching jet at time $t_{\rm{ya}}$ in
  order to reach plane $f$ at time $t_{\rm{f}}$.  Light from position
  $y$ within the corresponding plasmon in the receding jet leaves at
  an earlier time $t_{\rm{yr}}$ in order to reach $f$ at this same
  time.  All light leaving $f$ at time $t_{\rm{f}}$ will obviously be
  received at the same time by the observer. \label{fig:jet}}
\end{figure}

Our toy model is shown in Figure\,\ref{fig:jet}; our analysis follows
that of \citet{Blu94} who considered analogous effects in the
evolution of the lobes of radio galaxies.  We consider light setting
off from the closest point of the first plasmon to the observer, and
determine what time the light would have left other parts of that
plasmon, and other plasmons, in order to reach plane $f$ at the same
time.  All light leaving plane $f$ (perpendicular to the line of
sight) at a given instant clearly reaches the observer simultaneously.
The refractive index of the plasmons is taken to be 1 (in reality, it
will be greater than 1, and this will make the light travel-time
effects even more significant).  We assume a tangled field, so that
the magnetic flux density falls off as $B\propto R^{-1}$ (as described
in \S\,\ref{sec:ssa}).  Given the magnetic field at any one time, and
having calculated the time the light set off from a plane within a
plasmon, and knowing the time at which the plasmon was ejected, then
assuming constant uniform expansion of the plasmons, the size of a
plasmon at any given time, and therefore its magnetic field can be
determined.  Dividing each plasmon up into slices perpendicular to the
line of sight, from which light sets off progressively earlier, we
derive the field in each slice at the time at which light leaves that
slice.  We can then calculate the emissivity of each slice, assuming
the plasmons to be homogeneously filled with energetic particles.  We
then integrate over the entire plasmon to find the flux density
expected at any particular frequency, having corrected for Doppler
boosting and relativistic time-dilation effects.

To model the emissivity of each slice, rather than take the full
expression for the synchrotron emissivity, we assume that each
electron radiates at a single frequency, determined by its Lorentz
factor and the ambient magnetic field (Equation\,\ref{eq:nu_gamma}).  In order
to model the opacity, we put in an absorption coefficient (derived
either for synchrotron self-absorption, or for free-free absorption,
as in \citet{Ryb79}).  We calculate the location of the $\tau=1$
surface within each plasmon at each frequency, and assume we see no
emission from beyond that surface.  Emission from the near side of the
$\tau=1$ surface is taken to be synchrotron emission from a power-law
distribution of electrons with electron index $p=2.2$.  This produces
a turned-over spectrum as we observe.  We then observe how the spectrum
evolves with time.

\begin{figure}
\psfig{figure=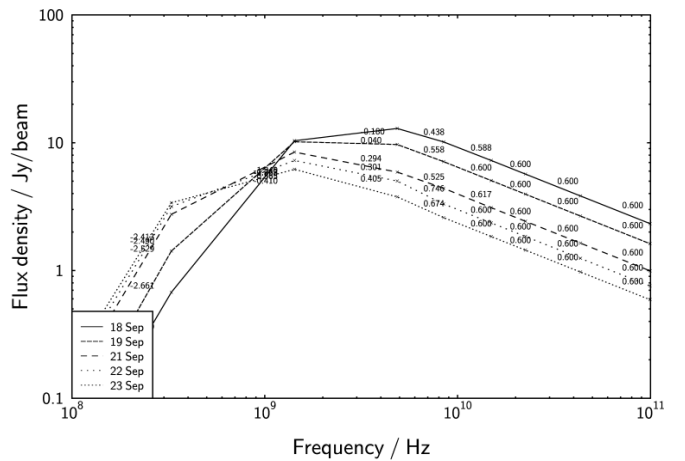,width=8.8cm,angle=0,clip=t}
\caption{Model spectra for 2001 September 18, 19, 21, 22, and 23, to
  be compared with Figure\,\ref{fig:spectra}.  Spectral indices
  between each sampled frequency point are given.  The model takes
  into account light travel-time and opacity effects for a series of
  jet knots, with adiabatic expansion of a series of ejected plasmons
  governing the spectral evolution.  As shown above, it does not
  include any flat-spectrum core component.
  \label{fig:model_spectra}}
\end{figure}

Plots of spectra generated are given in
Figure\,\ref{fig:model_spectra}.  The full theory will be explored in
more detail in a companion paper (Miller-Jones et al.\ in prep.).  For
now, we note that such a model produces the qualitative features of
spectra similar to those observed, in that it produces a turnover
moving to lower frequencies with time and a steepening of the spectrum
at frequencies just above the turnover to a terminal value of the
spectral index which corresponds to the underlying electron energy
index.  An important success of this model is that the timescales for
the movement of the turnover and for the steepening of the
high-frequency spectrum match those observed.

However, where this model falls short is that it does not predict a
spectrum flatter than $\nu^{-0.6}$ all the way up to the highest
observing frequencies.  However, this model only includes jet knots
which evolve as described above: it takes no account of any
flat-spectrum core component.  We note that invoking the addition of a
core with a canonical $\nu^{-0.1}$ spectrum whose intensity decreases
with time might sufficiently flatten the predicted high-frequency
spectrum at early epochs.  Such a core component was postulated in
\S\,\ref{sec:vlba_spectra} (see also Figure\,\ref{fig:precession}) and
was present in this outburst, as we now describe.  In
\S\,\ref{sec:flare_start}, we found that the observed VLBA knots
appeared to have been ejected $2.5\pm0.5$ days after the integrated
radio flux density began to rise.  This would be consistent with a
scenario in which a flat-spectrum core component as postulated above
dominated the flux density initially, but faded over time.  After the
ejection of the steep-spectrum radio plasmons, their flux density
would later come to dominate the observed flux density, especially at
lower frequencies.

\section{Comparison to radio galaxies}\label{sec:big_guys}
We now make a brief comparison between the magnetic field and energy
budget in this microquasar flare and those parameters in quasars:
scaled-up, longer-lived versions of \cyg.

\subsection{Magnetic Field}

M\,87 is a radio galaxy at a distance of 16\,Mpc for which there are
good constraints on the magnetic field at different distances from the
core, which has a central black hole of mass $3.2\pm 0.9\times
10^9$\,\msun \citep{Mac97}.  \citet{Rey96} derived limits to the
magnetic field in the jet of M\,87 of $0.01<B<0.2$\,G at an angular
separation of 0.1\,mas from the core, approximately 1800\,AU, or 40
gravitational radii ($r_{\rm{G}}$) from the central engine.
\citet{Hei97} derived a magnetic field of $5<B<49\,\mu$G at several
kpc ($\sim 10^6-10^7 r_{\rm{G}}$) from the core.  Standard
equipartition arguments give $300<B<600\,\mu$G at that distance
\citep[e.g.][]{Mei96}.

\cyg\ is thought to contain either a neutron star or a black hole of a
few solar masses \citep[e.g.][]{vandenHeu73, Sch96, Erg98, Sta03}: a
factor of $10^8-10^9$ times smaller than the black hole in M\,87.
Assuming that the compact object in \cyg\ has a mass of a few solar
masses, its gravitational radius is of order 10\,km.  Observing with a
few mas resolution probes the source at $10^8-10^9\,r_{\rm{G}}$.
According to \citet{Hei02}, magnetic fields in the inner jet and disk
should scale as $M^{-1/2}$ where $M$ is the black hole mass.
Comparing the magnetic fields in the inner jets of M\,87 and \cyg\,
this predicts the field in the inner jet (tens of gravitational radii
from the compact object) of \cyg\ to be of order $10^3$\,G.  This is a
factor of $10^4$ higher than that which we observe further out at
$\sim 10^8\,r_{\rm{G}}$.  Let $R$ be the jet cross-section and $r$ the
distance from the core measured along the jet axis.  Taking a jet
model in which the cross-section scales as $R\propto r^{\delta}$, and
invoking flux-freezing and a tangled magnetic field, such that
$B\propto R^{-1}$, we find that $B\propto r^{-\delta}$.  Putting in
the values we have for $B$ and $r$, we find that $0.50<\delta<0.83$.
This limit may be compared with theoretical limits quoted by
\citet{Hei02b}.  They found that
\begin{alignat}{2}
R &= R_0(r/r_0)^{\beta}, \qquad &0<\beta<1 \\
n &= n_0(r/r_0)^{-\lambda}, \qquad &2\beta\leq\lambda\leq3\beta \\
B&=B_0(r/r_0)^{-\delta}, \qquad &1/2\lambda\leq\delta\leq\lambda
\end{alignat}
where they defined $r$ as the distance from the black hole, $r_0$ as
the distance at which the jet is injected (taken as 10\,$r_{\rm G}$),
$R$ as the jet cross-section and $n$ as the particle number density.
They did however point out that this assumes self-similarity.  It
would thus be expected to hold in the inner jet, but not necessarily
further out where the jet behaviour may be dominated by interactions
with its environment.  The case $\beta = 1$ corresponds to a jet with
constant opening angle.  A jet composed of discrete ejections has
$\lambda\sim 3\beta$, whereas a continuous jet would satisfy
$\lambda\sim 2\beta$.  A value $\delta = 2/3\lambda$ is predicted for
a completely tangled magnetic field.  From the VLBA observations in
Section\,\ref{sec:vlba}, we observed a jet which seems to be composed
of discrete ejections, and we have assumed a tangled field.  Thus we
would expect $\delta=2\beta$.  Our derived limits on the value of the
index $\delta$ then imply $0.25<\beta<0.41$.  Since the VLBA images
did not have sufficient transverse resolution to monitor the jet
opening angle, and since they showed a series of discrete knots in the
jet, it is not possible to verify accurately any variation in the
opening angle from the observations.

We note, however, that although we expect similar physics to apply in
the inner, jet-producing regions of AGN and microquasars, the
environments of the two types of system are different.  According to
\citet{Hei02}, in dynamical terms, microquasar jets are located in
relatively low-pressure, low-density environments.  Where we observe
the radio jets (i.e.\ as they interact with their environments), the
characteristics of the environment will affect the kinematics of the
jet, and a simple scaling between the two types of system is not
expected.

\citet{Hei97} found that the magnetic field in the M\,87 jet at a
distance of several kpc from the core should be a factor $\sim1.5-5$
times lower than the equipartition value.  We found the Cygnus X-3
outer jet to be out of equipartition (modulo the uncertainties in this
calculation) by a similar factor of a few (\S\,\ref{sec:num_density}).

\subsection{Energetics}
The typical energy thought to be stored in the lobes of a radio galaxy
such as Cygnus\,A is estimated to be of order $10^{59}$\,erg
\citep[][chapter 9]{Kem99}, and the typical kinetic jet power of order
$10^{42}-10^{43}$\,erg\,s$^{-1}$ \citep{Rey96}.  From the results of
September 18, it is possible to deduce the minimum energy present in
the outburst, using the $\gamma_{\rm{max}}$ and $\gamma_{\rm{min}}$
corresponding to the highest and lowest observed frequencies on that
date.  This gives a lower limit of $8.5 \times 10^{42}$\,erg for the
energy released following this flare.  We do not believe, however,
that we have fully sampled the range of Lorentz factors present in the
jet, so this should be taken as a lower limit, and fuller sampling
would be required in order to make a complete comparison with AGN.

\section{Conclusions}

We have presented time-sequenced milliarcsecond-resolution
observations at three different frequencies of a major outburst of
\cyg, accompanied by simultaneous quasi-daily monitoring of the
(unresolved) flux density of the source, with unprecedented spectral
coverage of a large flare, ranging over three decades in frequency.
The proper motions of individual plasmons have been directly tracked
for the first time, and the expansion of those plasmons has been
directly measured.

One main result is to securely determine the presence of a two-sided
jet in \cyg\ on milliarcsecond scales following a major flare.  The
southern jet appears to be approaching and the northern counterjet
appears to be receding from us, being substantially de-boosted.  We
derived a jet speed of $\beta\sim0.63$.  The jet appears to be precessing
with a period of order 5.3\,days, significantly shorter than
that derived by \citet{Mio01} for the 1997 flare.  It appears that the
milliarcsecond-scale jet is not ejected immediately on the rise of the
integrated flux density, but $\sim 2.5$ days after this, a point which
we explore in a forthcoming paper (Miller-Jones et al.\ in prep.).

We believe that it is possible that the intrinsic
characteristics of \cyg\ in the outbursts of, for example, 1997 and
2001, may have been different.  It is not clear that all outbursts of
such sources follow the same patterns.  Certainly the flux densities
and fading times are not the same for different events, as outlined
above.  This variable, almost whimsical, nature of the source needs to
be clarified.  Certain parameters would not be expected to change
(e.g.\ after accounting for precession, the inclination angle of the jet
axis to the line of sight might be expected to be constant).
Identical spectral, temporal, and \textit{uv}-coverage of different
events with similar radio lightcurves might help to determine whether
the jet parameters are indeed constant for different flares.

Observations of the integrated flux density of the source showed an
evolving spectrum similar to that seen in previous outbursts of
\cyg. The high-frequency spectrum at frequency $\nu$ took the form
$\nu^{-\alpha}$, where $\alpha$ steepened from $\sim 0.4$ to $\sim 0.6$
and the spectral index then remained at this latter terminal value.
We contend that the spectral evolution is the result of light
travel-time effects in plasmons which were initially optically thick,
but which became optically thin as they expanded.  Such a model can
produce spectral evolution over timescales similar to those observed.
Confirmation that opacity effects are important came from the
integrated spectrum between 1.4\,GHz and 74\,MHz: this showed a
turnover moving to lower frequencies with time, caused by the
decreasing opacity (probably caused by either synchrotron
self-absorption, free-free absorption or a low-energy cutoff in the
electron spectrum).  The presence of an additional flat-spectrum core
component which was responsible for the initial rise of the flare but
which then faded with time was also postulated.

We derived an upper limit to the magnetic field in the plasmons of
$0.15\pm0.06$\,G, consistent with previous estimates for similar
systems.  We also placed lower limits on the minimum and maximum
observed Lorentz factors of the emitting electrons, of $\gamma=58$ and
$\gamma=321$.

Despite the unique nature of this system, \cyg\ does share
many of the characteristics of other, better understood microquasars.
Owing to its position within the Galaxy and the associated
difficulties in imaging the system, \cyg\ still remains somewhat of an
enigma.  Studying it can, however, enhance our understanding of the
effects which we believe must play a part in the observed emission
from microquasars.  Light travel-time effects are unavoidable in
systems of this size evolving on similar timescales.

\acknowledgments

The authors would like to thank Philipp Podsiadlowski, Sebastian Heinz
and Bryan Butler for useful discussions.  JCAM-J thanks the UK
Particle Physics and Astronomy Research Council for a Studentship.
KMB thanks the Royal Society for a University Research Fellowship.
The VLA and VLBA are facilities of the NRAO operated by Associated
Universities, Inc., under co-operative agreement with the National
Science Foundation.  Research with the Owens Valley Radio Observatory
Millimeter Array, operated by California Institute of Technology, is
supported by NSF grant AST 99-81546. AJB gratefully acknowledges the
support of NSF grant AST 01-16558.  KMB and PD acknowledge a joint
British Council/Enterprise Ireland exchange grant.  MPR thanks Oxford
University Astrophysics for support during his visit to Oxford, and
JCAM-J and KMB thank NRAO for their support during their visits to
Socorro.  The authors also thank the referee for a careful reading of
the manuscript.

\clearpage
\onecolumn

\newpage

\begin{deluxetable}{cccccccc}
\tabletypesize{\scriptsize}
\tablewidth{0pt}
\tablecolumns{4}
\tablecaption{\cyg\ VLA Flux Densities \label{tab:vla_observations}}
\tablehead{
\colhead{Date (UT)} & \colhead{JD-2450000}   & 
\colhead{Observing Frequency (GHz)} & \colhead{Flux Density (Jy)}
}
\startdata
2001 September 09 & $2161.73955\pm0.00050$ & \phn4.860 &
$0.0209\pm0.0007$ \\
 & $2161.73356\pm0.00046$ & \phn8.460 & \phn$0.0312\pm0.0005$ \\
 & $2161.73212\pm0.00075$ & 14.94\phn & \phn$0.0424\pm0.0020$ \\
 & $2161.72419\pm0.00127$ & 22.46\phn & \phn$0.0387\pm0.0030$ \\
 & $2161.71748\pm0.00139$ & 43.34\phn & \phn$0.0472\pm0.0038$ \\
\hline
2001 September 18 & $2170.57112\pm0.00168$ & \phn1.385 &
\phn$8.56\pm0.17$ \\
 & $2170.57112\pm0.00168$ & \phn1.465 & \phn$9.16\pm0.18$ \\
 & $2170.57570\pm0.00093$ & \phn4.860 & $13.41\pm0.27$ \\
 & $2170.57778\pm0.00093$ & \phn8.460 & $11.85\pm0.12$ \\
 & $2170.59215\pm0.00091$ & 14.94\phn & \phn$9.04\pm0.43$ \\
 & $2170.58810\pm0.00197$ & 22.46\phn & \phn$7.44\pm0.57$\tablenotemark{b} \\
 & $2170.58351\pm0.00168$ & 43.34\phn & \phn$5.99\pm0.46$ \\
\hline
2001 September 19 & $2171.45749\pm0.00050$ & \phn1.385 &
$14.25\pm0.28$ \\
 & $2171.45749\pm0.00050$ & \phn1.465 & $14.31\pm0.29$ \\
 & $2171.45660\pm0.00058$ & \phn4.860 & $13.43\pm0.27$ \\
 & $2171.46615\pm0.00052$ & \phn8.460 & $10.54\pm0.11$ \\
 & $2171.48380\pm0.00058$ & 14.94\phn & \phn$8.22\pm0.39$ \\
 & $2171.48079\pm0.00127$ & 22.46\phn & \phn$6.70\pm0.51$\tablenotemark{b} \\
 & $2171.47454\pm0.00127$ & 43.34\phn & \phn$4.62\pm0.37$ \\
\hline
2001 September 21 & $2173.63970\pm0.00058$ & \phn1.385 & \phn$7.98\pm0.16$
\\
 & $2173.63970\pm0.00058$ & \phn1.465 & \phn$7.86\pm0.16$ \\
 & $2173.64092\pm0.00041$ & \phn4.860 & \phn$4.83\pm0.10$ \\
 & $2173.65000\pm0.00046$ & \phn8.460 & \phn$3.54\pm0.04$ \\
 & $2173.66476\pm0.00041$ & 14.94\phn & \phn$2.37\pm0.11$ \\
 & $2173.66119\pm0.00050$ & 22.46\phn & \phn$1.79\pm0.14$ \\
 & $2173.65567\pm0.00046$ & 43.34\phn & \phn$1.22\pm0.15$ \\
\hline
2001 September 22 & $2174.52118\pm0.01806$ & \phn0.3275 &
\phn$6.161\pm1.4$\phn\phn \\
 & $2174.50232\pm0.00058$ & \phn1.385 & \phn$5.640\pm0.113$ \\
 & $2174.50232\pm0.00058$ & \phn1.465 & \phn$5.377\pm0.108$ \\
 & $2174.52558\pm0.01782$ & \phn4.860 & \phn$2.968\pm0.059$ \\
 & $2174.53115\pm0.01587$ & \phn8.460 & \phn$2.126\pm0.021$ \\
 & $2174.53671\pm0.00091$ & 14.94\phn & \phn$1.465\pm0.070$ \\
 & $2174.52646\pm0.00108$ & 43.34\phn & \phn$0.688\pm0.055$ \\
\hline
2001 September 23 & $2175.70995\pm0.01053$ & \phn0.0738 &
$<2.7$\tablenotemark{a} \\
 & $2175.70723\pm0.01048$ & \phn0.3275 & \phn$4.061\pm1.4$\phn\phn \\
 & $2175.69601\pm0.00052$ & \phn1.385 & \phn$2.919\pm0.058$ \\
 & $2175.69601\pm0.00052$ & \phn1.465 & \phn$2.834\pm0.057$ \\
 & $2175.70139\pm0.00046$ & \phn4.860 & \phn$1.425\pm0.029$ \\
 & $2175.70642\pm0.00041$ & \phn8.460 & \phn$0.990\pm0.010$ \\
 & $2175.71570\pm0.00062$ & 14.94\phn & \phn$0.641\pm0.031$ \\
 & $2175.71250\pm0.00058$ & 22.46\phn & \phn$0.492\pm0.038$ \\
 & $2175.71078\pm0.00091$ & 43.34\phn & \phn$0.315\pm0.030$ \\
\hline
2001 September 24 & $2176.69213\pm0.01563$ & \phn0.0738 &
$<4.5$\tablenotemark{a} \\
 & $2176.68918\pm0.01557$ & \phn0.3275 & $7.001\pm1.4$\phn\phn \\
 & $2176.69371\pm0.02157$ & \phn1.385 & $1.893\pm0.038$ \\
 & $2176.69371\pm0.02157$ & \phn1.465 & \phn$1.842\pm0.037$ \\
 & $2176.69618\pm0.01794$ & \phn4.860 & \phn$0.895\pm0.018$ \\
 & $2176.69699\pm0.03117$ & \phn8.460 & \phn$0.638\pm0.006$ \\
 & $2176.69884\pm0.00058$ & 14.94\phn & \phn$0.470\pm0.022$ \\
 & $2176.69220\pm0.00065$ & 22.46\phn & \phn$0.392\pm0.030$ \\
 & $2176.69041\pm0.00091$ & 43.34\phn & \phn$0.314\pm0.016$ \\
\enddata
\tablenotetext{a}{These are 3$\sigma$ upper limits as no source was
  detected at the known position of \cyg }
\tablenotetext{b}{Tipping scans were used to derive the atmospheric
  opacities for these observations}
\tablecomments{The error bars given were derived as described in
  \ref{sec:errors}}
\end{deluxetable}

\begin{deluxetable}{ccc}
\tabletypesize{\scriptsize}
\tablewidth{0pt}
\tablecolumns{3}
\tablecaption{\cyg\ OVRO Millimeter Array Flux Densities at 99.48\,GHz \label{tab:ovma_observations}}
\tablehead{
\colhead{Date (UT)} & \colhead{JD-2450000} & \colhead{Flux Density (Jy)}
}
\startdata
2001 Sep 20 & $2172.6930\pm0.0597$ & $1.206\pm0.005$ \\
2001 Sep 21 & $2173.6242\pm0.0167$ & $0.942\pm0.007$ \\
2001 Sep 22 & $2174.8232\pm0.0170$ & $0.363\pm0.010$ \\
2001 Sep 23 & $2175.6892\pm0.0205$ & $0.216\pm0.005$ \\
2001 Sep 26 & $2178.5559\pm0.0275$ & $0.121\pm0.004$ \\
\enddata
\tablecomments{The uncertainties quoted in the flux density are those
  due to thermal noise.}
\end{deluxetable}

\begin{deluxetable}{ccccccc}
\tabletypesize{\scriptsize}
\tablewidth{0pt}
\tablecolumns{7}
\tablecaption{VLBA image parameters \label{tab:vlba}}
\tablehead{
\colhead{Date (UT)} & \colhead{JD-2450000} & \colhead{$\nu$
  (GHz)} & \colhead{Beam (mas $\times$ mas)} & \colhead{Beam
  P.A.} & \colhead{Scattering Disk Size (mas)} & \colhead{r.m.s.\
  (mJy bm$^{-1}$)}
}
\startdata
2001 Sep 18 & $2170.78779\pm0.17781$  & 22.233 & \phn$8.59\times3.48$\phn & $-59.4$ & $\phn0.69\pm0.22$ & \phn19.7\phn \\
 & $2170.76849\pm0.18423$ & \phn4.995 & $13.71\times12.92$ & $+16.7$ & $15.54\pm2.42$ & \phn13.8\phn \\
\hline
2001 Sep 19 & $2171.61065\pm0.20914$ & 22.233 & \phn$8.30\times4.00$\phn & $-29.1$ & $\phn0.69\pm0.22$ &
\phn\phn9.68 \\
 & $2171.59074\pm0.20183$ & 15.365 & $16.19\times8.76$\phn & $-19.8$ & $\phn1.48\pm0.42$ & 126.4\phn \\
 & $2171.59149\pm0.21542$ & \phn4.995 & $15.65\times13.58$ & $-63.9$ & $15.54\pm2.42$ &
\phn\phn6.38 \\
\hline
2001 Sep 20 & $2172.75584\pm0.20548$ & 22.233 & \phn$5.88\times3.77$\phn & $-59.2$ & $\phn0.69\pm0.22$ &
\phn\phn6.52 \\
 & $2172.72633\pm0.18880$ & 15.365 & \phn$9.39\times8.91$\phn & $+19.2$ & $\phn1.48\pm0.42$ & \phn42.0\phn \\
 & $2172.71344\pm0.18880$ & \phn4.995 & $14.54\times13.45$ & $-85.8$ & $15.54\pm2.42$ & \phn17.9\phn \\
\hline
2001 Sep 21 & $2173.65190\pm0.26304$ & 15.365 & $13.31\times9.00$\phn & $-14.4$ & $\phn1.48\pm0.42$ &
\phn21.0\phn \\
 & $2173.63907\pm0.26305$ & \phn4.995 & $11.85\times10.60$ & $+61.4$ & $15.54\pm2.42$ & \phn\phn7.49 \\
\enddata
\end{deluxetable}

\begin{deluxetable}{cccccc}
\tabletypesize{\scriptsize}
\tablewidth{0pt}
\tablecolumns{6}
\tablecaption{Fitted Gaussian FWHM sizes of the 22\,GHz VLBA knots
  \label{tab:vlba_sizes}}
\tablehead{
\colhead{Date} & \colhead{Component} & \colhead{Fitted FWHM (mas)\tablenotemark{a}} &
\colhead{Major Axis (mas)} & \colhead{Minor Axis (mas)} & \colhead{P.A.}
}
\startdata
2001 Sep 18 & Single component & $7.88\pm0.51$ & $7.23\pm0.17$ & $2.65\pm0.34$ & $167.5\pm2.0$\\
2001 Sep 19 & North & $11.98\pm0.15$ & $11.00\pm0.14$ & $4.03\pm0.09$ & $168.1\pm0.7$\\
2001 Sep 19 & South & $13.36\pm0.21$ & $10.93\pm0.23$ & $5.04\pm0.12$ & $157.5\pm1.2$\\
2001 Sep 20 & North & $14.37\pm0.54$ & $10.58\pm0.39$ & $6.02\pm0.39$ & $26.5\pm94.3$\\
2001 Sep 20 & South & $16.92\pm0.88$ & $13.89\pm0.71$ & $6.36\pm0.58$ & $25.0\pm95.8$\\
\enddata
\tablenotetext{a}{The diameter of a sphere subtending the same solid
  angle as the Gaussian with the given major and minor axes,
  calculated according to Equation\,\ref{eq:theta}.}
\tablecomments{Deconvolved sizes and errors were derived using the
  \textit{AIPS} task JMFIT}
\end{deluxetable}

\begin{deluxetable}{cccccccc}
\tabletypesize{\scriptsize}
\tablewidth{0pt}
\tablecolumns{8}
\tablecaption{Results of fitting the 22\,GHz VLBA image of September
  20 with the Hjellming \& Johnston precessing jet model
  \label{tab:jet_fitting}}
\tablehead{
\colhead{Core R.A.} & \colhead{Core Dec.} & \colhead{$s_{\rm rot}$} &
\colhead{Parameter} & \colhead{Mean} & \colhead{Standard Deviation} &
\colhead{Maximum} & \colhead{Minimum} }
\startdata
20 32 25.77335 & +40 57 27.9650 & -1 & $\phi$ & 168.8 & 91.1 & 345.9 &
3.3 \\
 & & & $\psi$ & 2.4 & 1.2 & 5.8 & 0.4 \\
 & & & $i$ & 10.5 & 4.2 & 18.8 & 2.5 \\
 & & & $\chi$ & 98.9 & 2.9 & 106.2 & 93.4 \\
 & & & $\beta$ & 0.63 & 0.09 & 0.80 & 0.47 \\
 & & & $P$ (days) & 5.34 & 0.67 & 7.15 & 3.73 \\
 & & & $d$ (kpc) & 10.8 & 2.6 & 14.8 & 5.6 \\
 & & & $t_{\rm view}$ (days) & 7.44 & 0.54 & 8.43 & 6.08 \\
\hline
20 32 25.77335 & +40 57 27.9650 & +1 & $\phi$ & 136.1 & 97.7 & 330.2 &
2.1 \\
 & & & $\psi$ & 2.2 & 0.9 & 4.3 & 0.8 \\
 & & & $i$ & 169.6 & 3.8 & 175.6 & 161.3 \\
 & & & $\chi$ & 278.1 & 2.8 & 283.6 & 273.6 \\
 & & & $\beta$ & 0.60 & 0.08 & 0.77 & 0.49 \\
 & & & $P$ (days) & 5.49 & 0.46 & 7.06 & 4.81 \\
 & & & $d$ (kpc) & 10.0 & 2.1 & 13.0 & 6.2 \\
 & & & $t_{\rm view}$ (days) & 7.53 & 0.52 & 8.42 & 6.61 \\
\enddata
\tablecomments{All angles ($\phi$, $\psi$, $i$, and $\chi$) are
  measured in degrees}
\end{deluxetable}

\end{document}